\documentclass[lettersize,journal]{IEEEtran}
\usepackage[noadjust]{cite} % Improves presentation of citations, like [3--7].
\usepackage{multirow}
\usepackage{amsmath}
\usepackage{overpic}
\usepackage{amsopn}
\usepackage{amssymb}
\usepackage{accents}
\usepackage{mathrsfs}
\usepackage{pifont}

\usepackage{bm}         % for nice bold math symbols using \bm{...}
\usepackage{fix-cm}     % permit arbitrary sizes of Computer Modern

\usepackage{mleftright} %provides \mleft, \mright to fix issue with spacing
\mleftright                 %redefines all \left and \right to behave
\usepackage{mathdots}   % fixes \ddots and \vdots to scale properly

\makeatletter
\def\IEEElabelanchoreqn#1{\bgroup
\def\@currentlabel{\p@equation\theequation}\relax
\def\@currentHref{\@IEEEtheHrefequation}\label{#1}\relax
\Hy@raisedlink{\hyper@anchorstart{\@currentHref}}\relax
\Hy@raisedlink{\hyper@anchorend}\egroup}
\makeatother

\usepackage[utf8]{inputenc} 
\usepackage[T1]{fontenc} % To force 8-bit encoding so that
\usepackage{widetext}
\usepackage{tikz}
\usepackage{tikz}
\allowdisplaybreaks
\usetikzlibrary{datavisualization} 

\usetikzlibrary{positioning, fit, calc} % used for the efficient working of the positioning system
\tikzset{block/.style={draw, thick, minimum width=0.5cm, minimum height=0.5cm, align=center}, % the align command is used to align the block diagram at the center
	line/.style={-latex}   % the lesser the width the greater will be the diagram window
}

\usepackage{pgfplots}

\usepackage{subfigure}
\usepackage{algorithmicx}
\usepackage{algorithm}
\usepackage{algpseudocode}
\usepackage{amsthm}
\usepackage{epstopdf}
\usepackage{booktabs}
\usepackage{balance}

\usepgfplotslibrary{units}
\usetikzlibrary{spy,backgrounds}
\usetikzlibrary{decorations.markings}
\usepgflibrary{arrows.meta}
\usetikzlibrary{positioning}
\usetikzlibrary{shadows} %used also in mdframed
\usetikzlibrary{calc}
\usetikzlibrary{patterns}
\usepackage{pgfplots}
\pgfplotsset{compat=1.15}
\usepgfplotslibrary{units}
\usetikzlibrary{spy,backgrounds}
\usetikzlibrary{decorations.markings}
\usepackage{pgfplotstable}

\newtheorem{theorem}{Theorem}

\newtheorem{corollary}{Corollary}

\theoremstyle{remark}

\usepackage{hyphenat} % Package for hypenation-stuff

\newcommand{\midk}[1]{\kern0.1em #1 \kern0.1em}
\newcommand{\middlek}[1]{\kern0.1em \middle#1 \kern0.1em}
\newcommand{\bigk}[1]{\kern-0.1em \bigm#1 \kern-0.1em}
\newcommand{\Bigk}[1]{\kern-0.1em \Bigm#1 \kern-0.1em}
\newcommand{\biggk}[1]{\kern-0.1em \biggm#1 \kern-0.1em}
\newcommand{\Biggk}[1]{\kern-0.1em \Biggm#1 \kern-0.1em}

\newcommand{\const}[1]{\textnormal{\usefont{U}{eur}{m}{n}\selectfont #1}} % Euler
\newcommand{\hh}{\mathop{}\!\const{h}}  % differential entropy
\newcommand{\II}{\mathop{}\!\const{I}}  % mutual information
\newcommand{\relD}{\mathop{}\!\mathsf{D}}         % relative entropy
\newcommand{\relDf}[2]{\relD\left(#1 \kern0.1em\middle\|\kern0.1em #2\right)}
\newcommand{\erelDf}[2]{\relD(#1 \kern0.1em\|\kern0.1em #2)} 
\newcommand{\bigrelDf}[2]{\relD\bigl(#1 \kern-0.1em \bigm\| \kern-0.1em#2\bigr)}
\newcommand{\BigrelDf}[2]{\relD\Bigl(#1 \kern-0.1em \Bigm\| \kern-0.1em#2\Bigr)}
\newcommand{\biggrelDf}[2]{\relD\biggl(#1 \kern-0.1em \biggm\| \kern-0.1em#2\biggr)}
\newcommand{\BiggrelDf}[2]{\relD\Biggl(#1 \kern-0.1em \Biggm\| \kern-0.1em#2\Biggr)}

\newcommand{\Prvcond}[2]{\Pr\left[#1 \kern0.1em\middle|\kern0.1em #2\right]}
\newcommand{\ePrvcond}[2]{\Pr[#1 \kern0.1em|\kern0.1em #2]} 
\newcommand{\bigPrvcond}[2]{\Pr\bigl[#1 \kern-0.1em \bigm| \kern-0.1em#2\bigr]}
\newcommand{\BigPrvcond}[2]{\Pr\Bigl[#1 \kern-0.1em \Bigm| \kern-0.1em#2\Bigr]}
\newcommand{\biggPrvcond}[2]{\Pr\biggl[#1 \kern-0.1em \biggm| \kern-0.1em#2\biggr]}
\newcommand{\BiggPrvcond}[2]{\Pr\Biggl[#1 \kern-0.1em \Biggm| \kern-0.1em#2\Biggr]}

\newcommand{\Prscond}[2]{\Pr\left(#1 \kern0.1em\middle|\kern0.1em #2\right)}
\newcommand{\ePrscond}[2]{\Pr(#1 \kern0.1em|\kern0.1em #2)} 
\newcommand{\bigPrscond}[2]{\Pr\bigl(#1 \kern-0.1em \bigm| \kern-0.1em#2\bigr)}
\newcommand{\BigPrscond}[2]{\Pr\Bigl(#1 \kern-0.1em \Bigm| \kern-0.1em#2\Bigr)}
\newcommand{\biggPrscond}[2]{\Pr\biggl(#1 \kern-0.1em \biggm| \kern-0.1em#2\biggr)}
\newcommand{\BiggPrscond}[2]{\Pr\Biggl(#1 \kern-0.1em \Biggm| \kern-0.1em#2\Biggr)}

\newcommand{\Exp}{\operatorname{\textnormal{\textsf{E}}}}

\newcommand{\Econd}[3][]{\Exp_{#1}\left[#2 \kern0.1em\middle|\kern0.1em #3\right]}
\newcommand{\eEcond}[3][]{\Exp_{#1}[#2 \kern0.1em|\kern0.1em #3]}
\newcommand{\bigEcond}[3][]{\Exp_{#1}\bigl[#2 \kern-0.1em \bigm| \kern-0.1em #3\bigr]}
\newcommand{\BigEcond}[3][]{\Exp_{#1}\Bigl[#2 \kern-0.1em \Bigm| \kern-0.1em #3\Bigr]}
\newcommand{\biggEcond}[3][]{\Exp_{#1}\biggl[#2 \kern-0.1em \biggm| \kern-0.1em #3\biggr]}
\newcommand{\BiggEcond}[3][]{\Exp_{#1}\Biggl[#2 \kern-0.1em \Biggm| \kern-0.1em #3\Biggr]}

\newcommand{\dd}{\mathop{}\!\mathrm{d}}

\newcommand{\argmin}{\operatorname*{argmin}}
\newcommand{\argmax}{\operatorname*{argmax}}

\makeatletter

\newcommand{\Rmnum}[1]{\expandafter\@slowromancap\romannumeral #1@}
\makeatother

\usepackage{threeparttable}
\usepackage{diagbox}
\usepackage{array}
\usepackage{boldline}
\usepackage{caption}
\usepackage{dsfont} % less troublesome bbm fonts
\usepackage{algorithmicx}
\usepackage{algpseudocode}

\allowdisplaybreaks
\usetikzlibrary{datavisualization} 

\usetikzlibrary{spy,backgrounds}
\usetikzlibrary{decorations.markings}

\newcommand{\vol}{\operatorname{vol}}

\newcommand{\VV}{\const{V}}

\newcommand{\GG}{\const{G}}

\newcommand{\EEE}{\alpha}

\newcommand{\PP}{\const{P}}

\newcommand{\PAPR}[1]{\mathtt{PAPR} \left( #1 \right)}
\begin{document}

\title{Geometrically-Shaped Constellation for Visible Light Communications at Short Blocklength}

\author{Jia-Ning Guo, Ru-Han Chen{*}, Jian Zhang, Longguang Li, Xu Yang, and Jing Zhou%
 \thanks{This work is supported by Research Program of National University of Defense Technology under Grant No. ZK23-57 and ZK22-44, and the National Natural Science Foundation of China under Grant No. 62071489. Jia-Ning Guo and Jian Zhang are with National Digital Switching System Engineering and Technological Research Center, Henan Province (450000), China (e-mail: 14291003@bjtu.edu.cn, zhang\_xinda@126.com). Ru-Han Chen and Xu Yang is with Sixty-Third Research Institute, National University of Defense Technology, Nanjing, China (e-mail:tx\_rhc22@nudt.edu.cn, fractal\_yangxu@outlook.com). Longguang Li is with Dept. Communication and Electronic Engineering, East China Normal University, Shanghai, China (e-mail: lgli@cee.ecnu.edu.cn). Jing Zhou is with Dept. Computer Science and Engineering, Shaoxing University, Shaoxing, China (e-mail:jzhou@usx.edu.cn). \textit{(Corresponding Author: Ru-Han Chen.)}}}

\maketitle
\begin{abstract}
In this paper, we present a general framework of designing geometrically shaped constellations for short-packet visible light communications with a peak- and an average-intensity constraints. By leveraging tools from large deviation theory, we first characterize the second-order asymptotics of the optimal constellation shaping region under aforementioned intensity constraints, which serves as a good performance measure for the best geometric shaping in finite blocklength.
To further incorporate a sufficiently large coding gain and a nearly-maximum shaping gain, we construct multidimensional constellations by the nested structure of Construction B lattices, where the constellation shaping is implemented by controlling the boundary of the embedded sublattice, i.e., a strategy called coarsely shaping and finely coding. Fast algorithms for constellation mapping and demodulation are presented as well. As an illustrative example, we present an energy-efficient $24$-dimensional constellation design based on the Leech lattice, whose superiority over existing constellation designs is verified by numerical results.
	
\end{abstract}

\section{Introduction}
\label{Sec.introduction}
\IEEEPARstart{A}{s} a potential candidate for the next-generation wireless communication technology, visible light communication (VLC) has gained significant attention owing to integrated usage of communication and illumination, license-free deployment, inherent security, and absence of electromagnetic interference. Especially for less complexity and lower cost, intensity modulation and direct detection (IM/DD) are widely used in the VLC \cite{Zhang2019Intrinsic,Zhu2022Novel,Kafizov2022Probabilistic}, where the IM signal is required to be real and nonnegative since the information is modulated on the optical intensity emitted from light emitting diodes (LEDs). Moreover, a peak- and/or an average-intensity constraint may be imposed on the channel input, e.g., imposing an amplitude constraint usually for safety reason or suppressing the transceiver nonlinearity \cite{Mostafa2016Optimal}, an average-intensity constraint for dimming control \cite{Wang2013Tight}, or individual average-intensity constraints on multi-color LEDs for color adjustment \cite{Loureiro2024Multi}.
Those limitations lead to a fundamental difference between signaling for the IM signal in the VLC and the conventional electrical signal.

In indoor VLC scenarios, the main corruption brought by the strong background radiation and the thermal noise at the receiver photo diode (PD) can be modeled as the additive white Gaussian noise (AWGN), with which the IM/DD VLC channel is also known as \textit{the optical intensity channel}~\cite{kahnbarry97_1, hranilovic05_1,Chaaban2018Capacity,Li2020Capacity,Chen2022Capacity,Chaaban2022Capacity}. 
For this reason, at the birth of the VLC, the modulation and coding schemes are simply modified versions of those for radio-frequency (RF) or fiber communications. 
% In resent years, extensive research efforts have been conducted to characterize the capacity of the VLC under different constraints, which sheds light on advanced signaling designs for practical VLC systems \cite{Lapidoth2009Capacity,Farid2009Channel,Moser2018Capacity,Li2020Capacity,Chen2022Capacity}.
In recent years, to improve the system throughput, extensive research efforts have been conducted to advanced signaling designs for the practical VLC systems under different constraints. %\cite{Karout2012Two-Dimensional,Zhang2016Bandlimited,Zhang2018Block,Shiu1999Shaping,Zhang2016Energy,Farid2009Channel}.
In \cite{Karout2012Two-Dimensional,Zhang2016Bandlimited}, two-dimensional continuous-time signal spaces as well as constellation designs with respect to the corresponding discrete-time signal model are developed for the bandlimited VLC.
%\gjn{by maximizing the minimum Euclidean distance (MED) between constellation points,} 
%\gjn{by cooperative management of non-negative multiuser interference}
%\gjn{by adopting the nonequiprobable signaling technique for RF communication in \cite{Calderbank1990Nonequiprobable}}
%\gjn{by minimizing a total optical power subject to a fixed MED,}
For the VLC with a peak-intensity constraint, an efficient coded modem design is provided in \cite{Zhang2018Block}, of which the used constellation is a variant of lattice codes based on Construction A \cite[p. 29]{Zamir2014Lattice}. For the average-intensity limited VLC, probabilistically-shaped code and geometrically-shaped code are considered in \cite{Shiu1999Shaping} and \cite{Zhang2016Energy} respectively. Signaling under dual intensity constraints is more involved due to the need of more complicated constellation shaping. In \cite{Farid2009Channel}, a capacity-approaching non-uniform optical intensity signaling scheme under the dual intensity constraints is constructed by numerically optimizing the input distribution, while the channel coding is implemented by concatenating low density parity check (LDPC) codes with multilevel coding.

However, the above schemes either have notable performance loss due to lack of joint coding and shaping, or rely on long blocklengths for a larger coding gain or distribution mapping, and therefore, are not suitable for the short-packet VLC, which has various promising applications in industrial internet of things and vehicular communications \cite{Kinani2018Non-Stationary,Yang2020Coordinated,Zhou2023Design}.
Since short packets are utilized to carry critical control
information for ultra-low latency (e.g., smaller than 0.1 ms \cite{Niarchou2021Visible}), the blocklength of used channel codes is required to be very short \cite{Chang2023Low-Latency}. 
For the single-carrier VLC with a flat single-side bandwidth $B$ and using a strictly bandlimited and nonnegative Nyquist pulse, the duration time of $n$ successive symbols can be roughly evaluated by $\frac{n}{B}$\footnote{In \cite{Hranilovic2007Optical}, it is proved that the maximum ISI-free symbol rate of strictly bandlimited and nonnegative continuous-time waveforms is a half of that of electrical signals.}. Exemplified by a commercial white LED with $5$ MHz modulation bandwidth, to ensure that the user plane latency is within $0.1$ ms, it is required that the maximum blocklength can not exceed $500$. Taking the packet overhead, the probability of packet loss, and potential frequency multiplexing into consideration, the actual blocklength used in the packet payload will be far less than the above value.

%S. Chang, N. Huang, C. Gong and X. -Y. Li, "Low-Latency Network Slicing for VLC-Based Industrial Internet of Things: Superframe Duration Minimization and Delay Violation Probability Analysis," in IEEE Internet of Things Journal, vol. 10, no. 18, pp. 16617-16636, 15 Sept.15, 2023.

Motivated by the above fact, in this paper we are devoted to the signaling design for the short-packet VLC under a peak- and an average-intensity constraints\footnote{We would like to point out that another motivation for finite-blocklength analysis of constellation shaping comes from the fact that significant coding gains and shaping gains can be attained in relatively short blocklengths by lattice codes \cite{Lang2002A,Zamir2014Lattice}.}.
It is noted that there is no such known signaling design that approaches the Shannon limit (with respect to infinitely long blocklength) of the high signal-to-noise ratio (SNR) VLC under dual constraints at an acceptable cost, let alone characterizing the finite-blocklength limit. The main challenges in signaling for the short-packet VLC channel may come from two aspects:
\begin{itemize}
	\item \textit{Shaping:} 
 % Constellation shaping is widely acknowledged as to be necessary to approach channel capacity in a high-SNR regime~\cite{Forney1998Modulation}. 
	For the high-SNR VLC channel, directly using the amplitude shift keying (ASK) constellation even concatenated with the best channel code will remain a constant gap to the channel capacity, for example, 1.33 dB optical SNR loss (asymptotically) in the VLC channel with limited average intensity \cite{Shiu1999Shaping}, which reveals the necessity of the constellation shaping. Recently, geometric shaping has shown to be more efficient than probabilistic shaping in the short-blocklength regime, since the validity of distribution matching relies on a fairly long blocklength \cite{Qu2019Probabilistic,Amari2019Introducing}.
 %Unlike the AWGN channel with
	%only a quadratic constraint, the presence of amplitude
	%limitation usually leads to an irregular boundary of the
	%optimal shaping region, and hence, hinders involved
	%parameter selection and performance analysis in finite
	%dimensions [22]. 
    The main challenge brought by the finite-dimensional geometric
	shaping is to quantitatively measure the maximum shaping gain and quickly calculate corresponding parameters in the finite-blocklength regime.
	
	\item \textit{Coding:} In \cite{Chen2020FFT}, the authors construct an optimally-shaped constellation based on the checkerboard lattice $D_n$, which limits the nominal coding gain of the constructed constellation to only $1.5$ dB. To further approach the channel capacity, a more densely packing structure of constellation points should be utilized. 
%	\gjn{In \cite{Chen2020FFT}, for the VLC under the dual intensity constraints, the authors proposed a truncated cubic constellation (TCC) based on the checkerboard lattice $D_n$. The TCC scheme achieves a nominal coding gain of $1.5$ dB as compared with the ASK constellation. To overcome the limitation of the $D_n$ lattice on the coding gain, a more densely-packed lattice structure should be utilized.} 
	However, how to uniquely map the message onto the signal uniformly distributed within the optimal region, and efficiently incorporate the nearly-maximum shaping gain brought by an irregular region and a large coding gain by a densely-packed lattice in a coded modulation scheme are challenging.
 %In \cite{Chen2020FFT}, for the VLC under a peak- and an average-intensity constraints, the authors characterize the optimal geometrical shaping region and propose a low-complexity encoding algorithm that maps the message set onto the checkerboard lattice points enclosing by the optimal shaping region. 
\end{itemize}

In this paper, for the short-packet VLC with a peak- and an average-intensity constraints, we characterize the second-order asymptotics of the maximum shaping gain (i.e., analogous to the channel dispersion \cite{Hayashi2009Information, Polyanskiy2010Channel}), which touches upon the following important question in the short-packet VLC: \textit{How much shaping gain can be attained by low-dimensional geometric shaping?} To incorporate the nearly-maximum shaping gain and a significant coding gain, we propose a general framework for signaling in finite dimensions based on Construction B lattices. The main contributions are summarized as follows:
\begin{enumerate}
 \item \textit{Second-Order Asymptotics of Optimal Shaping:}
 Based on the large deviation theory, we characterize the second-order asymptotic behavior of the optimal shaping region, by which simple formulas for approximating the maximum shaping gain and the shaping parameter in the short-blocklength regime are therefore given. Numerical simulation reveals a good match of our proposed approximation even at low dimensions.

\item \textit{A General Signaling Framework:} Via exploiting the algebraic structure of the Construction B lattice, we propose a novel constellation design for the short-packet VLC, which not only is compatible with a flexible choice of blocklength, but also incorporates the nearly-maximum shaping gain and a significantly larger coding gain. We further present an illustrative example by using $24$-dimensional Leech lattice, which shows a performance gain of about $3$ dB over the conventionally-used ASK constellation.
\end{enumerate}

The remainder of this paper is organized as follows. The channel model is presented in Section~\ref{Sec.channel-model}. Some preliminaries are provided in Section~\ref{Sec.preliminaies}. In Section~\ref{Sec.shaping}, we characterize the second-order asymptotical properties of the optimal shaping region for the VLC under dual intensity constraints. In Section~\ref{Sec.coding}, a general framework for the optimally-shaped constellation construction is presented. An illustrative example and relevant numerical results are given in Section~\ref{Sec.simulation}. Section~\ref{Sec.conclusion} concludes the whole paper.

\section{Channel Model}
\label{Sec.channel-model}
In this paper, we consider a single-input single-output (SISO) VLC link whose channel output over $n$ successive channel uses can be modeled by
\begin{flalign}\label{eq.channel model}
\mathbf{r}=\mathbf{x}+\mathbf{z},
\end{flalign}
where the $n$-dimensional real vector $\mathbf{x}=\left( x_1,\cdots,x_n \right)$ denotes the instaneously transmitted intensity signal equiprobably chosen from some multidimensional constellation $\mathcal{X}$, and the $n$-dimensional vector $\mathbf{z}$ denotes the channel noise that modeled as the AWGN with zero mean and variance $\sigma^2$ \cite{Chaaban2022Capacity}. 

Due to the high-SNR property of the indoor VLC channel~\cite{Komine2004Fundamental} (also numerically verified in Section \ref{Sec.simulation}), in this paper we mainly concern with the multidimensional constellation design for intensity-modulated signals. Denote the cardinality of the constellation $\mathcal{X}$ to be designed by $M\in \mathbb{N}_+$. Then \textit{the normalize rate} of the constellation $\mathcal{X}$ is $\beta=\log_2 M/n$ bits per channel use (bpcu). 

Because of the nonnegativity of intensity-modulated signals, the constellation $\mathcal{X}$ is required to satisfy
%and hence satisfies the nonnegativity constraint:
$$
\mathcal{X}\subseteq \mathbb{R}_+^n,
$$
where the set $\mathbb{R}_+^n$ denotes the nonnegative orthant consisting of all nonnegative vectors in Euclidean $n$-space. Moreover, for the reasons of illumination control, device limitation and safety requirement in VLCs, a peak- and an average-intensity constraints are imposed on the constellation $\mathcal{X}$ as well\footnote{Without loss of generality, the maximum allowed peak intensity and the channel coefficient in the single-input single-output channel \eqref{eq.channel model} can be simultaneously normalized to unity by scaling the standard deviation $\sigma$ of the noise.}
\begin{subequations}\label{eq.constraint}
	\begin{align}
	% 0 \le X  \le 1,\label{eq.constraint1}\\ 
	% \mathbb{E}[X]\le\EEE\label{eq.constraint2}; 
 	\max_{\mathbf{x}\in\mathcal{X}}\Vert\mathbf{x}\Vert_{\infty} \le 1,\label{eq.constraintA}\\ 
	\dfrac{1}{nM}\sum_{\mathbf{x}\in\mathcal{X}}\sum_{i=1}^n x_i\le\EEE\label{eq.constraintB},
	\end{align}
\end{subequations}
where the constant $\EEE \in \left(0,\frac{1}{2} \right)$ denotes the ratio of the maximum allowed average intensity to the maximum allowed peak intensity. 

For invariance to scaling, in this paper we define the optical signal-to-noise ratio (OSNR) as the ratio of the maximum allowed peak intensity of the input signal to the standard deviation $\sigma$ of the noise, i.e.,
\begin{flalign}\label{eq:snr}
    \textrm{OSNR}\triangleq \frac{1}{\sigma}.
\end{flalign}

Throughout the paper, we assume that perfect knowledge of channel state information is available at the receiver, which is reasonable for the quasi-static VLC channel \cite{Farid2009Channel}. Under this assumption, it is well-known that the error performance of the constellation $\mathcal{X}$ at high SNR is primarily determined by the MED of the constellation $\mathcal{X}$ as follows:
\begin{flalign}
	d_{\text{min}}\left( \mathcal{X}\right) = \min_{\mathbf{x}_1,\mathbf{x}_2 \in \mathcal{X} \atop  \mathbf{x}_1 \neq \mathbf{x}_2 } \left\| \mathbf{x}_1-\mathbf{x}_2 \right\|_2.
\end{flalign}
Therefore, the main task of this paper is to design a multi-dimensional  constellation with a given normalized rate $\beta$ and a MED as large as possible under the constraints \eqref{eq.constraint}.

\section{Preliminaries}\label{Sec.preliminaies}
For both clarity and readability of our result, in this section we first briefly introduce some concepts that are frequently used throughout the paper.

\subsection{Construction B Lattice}
%\gjn{As a classical coded modulation scheme, the lattice codes can effectively integrate the coding gain and the shaping gain by selecting a dense set of lattice points in a given high-dimensional geometric region to design a constellation, which is widely used in bandlimited communication systems with high-SNR. In this paper, we design the  geometrically-shaped constellation based on the Construction B lattice.}
An $n$-dimensional \textit{lattice} is a subgroup of the Euclidean $n$-space with respect to the conventional vector addition operation. A \textit{Construction B} lattice is defined as
\begin{flalign}
H_n&=4\mathbb{Z}^n+2\left(n,n-1\right)+\mathcal{C},\nonumber\\
&=2D_n + \mathcal{C}
\label{eq.construction B}
\end{flalign}
where $\mathcal{C}$ represents an $\left(n,k_c,8 \right)$ binary linear block code and the notation $\left(n,n-1\right)$ refers to an $n$-dimensional parity check code, the checkerboard lattice $D_n$ is the sublattice of $\mathbb{Z}^n$ that consists of all points with even sum, i.e., 
$$D_n=\left\lbrace \left( x_1, \cdots, x_n \right) \in \mathbb{Z}^n : x_1+\cdots+x_n \ \text{is} \ \mathrm{even} \right\rbrace.$$ 
As illustrated in \cite{Forney1989bounded}, the MED of a Construction B lattice is $d_{\min}\left(H_n\right)=\sqrt{8}$.

\subsection{Truncated Cubes}
An $n$-dimensional \textit{truncated cube} $\mathscr{T}_{n} \left(t\right)$, with the largest coordinate $t\in [0,n]$, is the intersection of a unit $n$-cube with an $n$-simplex, i.e.,
\begin{flalign}\label{eq:truncated_cube}
	\mathscr{T}_n \left(t\right)\triangleq \left\{ \mathbf{x} \in [0,1]^n:\sum_{i=1}^{n}x_i \le t \right\},
\end{flalign}
whose volume and \textit{average first moment} are given by	
\begin{eqnarray}\label{eq.volume}
	\VV_n \left(  t\right)  =\frac{1}{n!}\sum_{k=0}^{n} \binom{n}{k}(-1)^{k} (t-k)^{n} \mathbf{1}_{\mathbb{R}_+}(t-k),
\end{eqnarray} 
and
\begin{flalign}\label{eq.moment}
	\PP_n \left(  t\right)&=
	\frac{1}{n	\VV_n \left(  t\right) }\int_{\mathscr{T}_n \left(t\right)} \left\|\mathbf{x} \right\|_1 \dd \mathbf{x}\\
	&=\frac{1}{n}\left(t-\frac{\sum_{k=0}^{n} \binom{n}{k}(-1)^{k} (t-k)^{n+1} \mathbf{1}_{\mathbb{R}_+}(t-k)}{\VV_n \left(  t\right)\cdot(n+1)!} \right),
\end{flalign}
respectively, where the indicator function $\mathbf{1}_{\mathbb{R}_+}(x)$ equals one if $x\ge 0$ and otherwise zero \cite{Chen2020FFT}.

\subsection{Geometric Shaping}
For an $n$-dimensional constellation $\mathcal{X}$ chosen from the lattice $\Lambda$, say the \textit{lattice code}, its \textit{shaping region} $\mathscr{R}$ is the closed region enclosing $\mathcal{X}$. In the following, we will present an information-theoretic definition of the shaping gain for the IM signal under a peak- and an average-intensity constraints, which is slightly different from its counterpart in coherent transmission \cite{Kschischang1994Optimal}.
\subsubsection{Shaping Gain}
Due to the amplitude constraint \eqref{eq.constraintA}, it is required that the shaping region satisfies $\mathscr{R}\subseteq [0,1]^{n}$ for IM signals. In line with   \cite{Zhang2018Block,Forney1989Multidimensional1}, the baseline shaping region is given as the one-dimensional line segment $\mathfrak{L}^{\dagger}=[0,2\alpha]$. In the high-SNR regime, we utilize the $n$-dimensional uniform distribution $\mathbf{W}_n \sim \text{Unif}\, \left( \mathscr{R} \right)$ and one-dimensional uniform distribution $W^{\dagger}\sim \text{Unif}\, \left( \mathfrak{L}^{\dagger}\right)$ as the channel input distribution respectively. Then the difference between their achievable rates can be approximated by
\begin{flalign}
&\dfrac{1}{n}\II\left( \mathbf{W}_n; \mathbf{W}_n+ \mathbf{Z} \right)-\II\left( W^{\dagger}; W^{\dagger}+ {Z} \right) \nonumber \\
\approx &\dfrac{1}{n} \log \left(  \vol\left(  \mathscr{R}  \right)  \right) - \log\left(  2\alpha  \right),
\end{flalign}
where ${\rm vol}\left( \mathscr{R}\right) $ is the $n$-dimensional volume of $\mathscr{R}$.
Due to the logarithm growth of the achievable rates for AWGN channels at high SNRs, we define \textit{the shaping gain} as
\begin{flalign}\label{eq.SGVLC}
\mathtt{SG}_{\textnormal{VLC}}\left(  \mathscr{R}  \right)
\triangleq
\exp   \left(   \frac{1}{n} \log \left(  \vol\left(  \mathscr{R}  \right)  \right) - \log\left(  2\alpha  \right)\right) 
=\frac{\sqrt[n]{ \vol\left(  \mathscr{R}  \right) }}{2\alpha},
\end{flalign}
which is also compatible with another definition in \cite{Shiu1999Shaping,Forney1989Multidimensional1,Chen2020FFT} that purely based on the high-dimensional geometry. 

\subsubsection{Optimal Shaping Region}
Due to the form of Eq. \eqref{eq.SGVLC}, we refer the \textit{optimal shaping region} to the closed region $\mathscr{R}$ with the maximized volume and satisfying corresponding constraints.
For the IM signal under a peak- and an average-intensity constraints \eqref{eq.constraint}, finding the optimal shaping region in $n$-space can be formulated as the following problem:
\begin{equation}\label{prob:opt_shaping}
	\begin{aligned}
		&\textnormal{max} &&\quad\quad \frac{\sqrt[n]{{\rm vol}\left( \mathscr{R}\right)}}{2 \EEE}\\
		&\textnormal{s.t.} &&\quad\quad \mathscr{R} \subseteq [0,1]^n\\
		& && \hspace{2em} \frac{1}{  \vol\left( \mathscr{R}\right)} \int_{\mathscr{R}} \left\| \mathbf{x} \right\|_1 \dd\mathbf{x}
		\le n\EEE,
	\end{aligned}
\end{equation}
where the last inequality is a continuous version of the average-intensity constraint \eqref{eq.constraintB} (see \cite{Forney1989Multidimensional1} for a detailed introduction of continuous approximation). 

In \cite{Chen2020FFT}, the optimal shaping region for the VLC channel under dual intensity constraints has been proved to be the family of truncated cubes. For self-containment, this result is reviewed in the following theorem.

\begin{theorem}(Optimal Shaping \cite{Chen2020FFT})\label{thm.shaping}
	The optimal solution to the problem \eqref{prob:opt_shaping} is $\mathscr{T}_n \left(t_n^{\star}\right)$ as defined in Eq. \eqref{eq:truncated_cube},  where the parameter $t_n^{\star}$ is determined by
	\begin{equation}\label{eq.shaping parameter}
		t_n^{\star}=
		\begin{cases}
			\PP_n^{{-1}}\left(\EEE \right) &,~\mbox{if  $\frac{1}{n+1} \le \EEE < \frac{1}{2}$},\\
			(n+1)\EEE &,~\mbox{if $\EEE \le \frac{1}{n+1}$},\\
		\end{cases}
	\end{equation}
	and $\PP_n(\cdot)$ is given by Eq. \eqref{eq.moment}. Accordingly, the maximum shaping gain in Euclidean-$n$ space is given by 
	\begin{equation}\label{eq.max_shaping_gain}
		\overline{\mathtt{SG}}_{\textnormal{VLC}}(n;\alpha)= \mathtt{SG}_{\textnormal{VLC}}\left(\mathscr{T}_n \left(t_n^{\star}\right)\right) =\sqrt[n]{\VV_n \left(t_n^{\star}\right)}/2\alpha.
	\end{equation} 
\end{theorem}

\subsection{$D_n$ Lattice Points in Truncated Cubes}
Next, we review the definition of the $D_n$-based truncated cube developed in \cite{Chen2020FFT}.
Let $H$ and $L$ be two nonnegative integers. 
%Then we define the \textit{$n$-dimensional discrete cube} with the parameter $H$   as $$\mathcal{C}_n(H)\triangleq \left\lbrace 0,1,\cdots,H \right\rbrace^n,$$
%The first moment shell $\mathcal{F}_n(H,l)$ with parameter $l$ is defined as:
%\begin{flalign}\label{eq.first moment shell}
%\mathcal{F}_n({H},\ell) \triangleq \left\{  \mathbf{x}\in \mathcal{C}_n(H)
%\,	| \, \sum_{i=1}^{n}x_i = \ell \right\},
%\end{flalign}
%where $\ell \in \left\lbrace 0,1,\cdots,nH \right\rbrace$. 
%The \textit{$D_n$-based cube} with the parameter $H$ as %$ \mathcal{DC}_n(H)$
%\begin{flalign}\label{eq.Dn based cube}
%	\mathcal{DC}_n(H)\triangleq D_n \cap \mathcal{C}_n(H),
%\end{flalign}
%i.e., the set of $D_n$ points in  $\mathcal{C}_n(H)$. 
The \textit{$D_n$-based truncated cube} $\mathcal{TD}_{n}\left({H},2{L}\right)$ with parameters $H$ and $L$ is given by
\begin{equation*}\label{eq.Dn based truncated cube}
	\mathcal{TD}_n\left({H},2{L} \right)
	\triangleq
	\left\{  \mathbf{x}\in D_n \cap \left\{0,1,\cdots,H\right\}^n :
	\sum_{i=1}^{n}x_i \le 2L \right\},
\end{equation*}
 i.e., the subset of the checkerboard lattice consisting of all points with the $\ell_1$-norm no larger than $2{L}$ and amplitudes between $0$ and $H$. We further let $\mathcal{TD}_{n}\left({H},2{L},M\right)$ denote the subset of $\mathcal{TD}_{n}\left(H,2{L}\right)$, which contains $M$ points with least $\ell_1$-norm in $\mathcal{TD}_{n}\left(H,2{L}\right)$.

\section{Optimal Shaping in Finite-Blocklength Regimes} \label{Sec.shaping}
Commonly-used shaping methods include probabilistic shaping and geometric shaping. The former one may not be suitable for the high-SNR case due to the need of distribution matching for an input set with a large cardinality \cite{Farid2009Channel}. Instead, a classic coded modulation scheme, called the \textit{lattice codes}, is widely used for bandlimited channels. The lattice code can be naturally combined with geometric shaping by the intersection of a densely-packed lattice (or its translate) and a well-chosen region in Euclidean space. In this section, based on the large deviation theory, we strength the result of geometric constellation shaping given in \cite{Chen2020FFT} (also see Theorem \ref{thm.shaping}) through the lens of finite-blocklength analysis.

\subsection{Finite-Blocklength Analysis of Optimal Shaping Parameter}

As shown in Theorem \ref{thm.shaping}, the optimal shaping parameter $t_n^{\star}$ should be numerically computed by the inverse of the average first moment $\PP_n(\cdot)$ that is strictly increasing. Although the bisection algorithm is applicable for solving \eqref{eq.shaping parameter}, the numerical method may become impractical as the blocklength increases, due to the presence of factorials and polynomials in \eqref{eq.volume} and \eqref{eq.moment}. Additionally, lack of an analytical evaluation for the optimal shaping parameter $t_n^{\star}$ may hinder a concrete analysis of the maximum shaping gain. To overcome this limitation, an asymptotical analysis of the optimal shaping region based on the large deviation theory is carried out in the following. 

Without loss of generality, we let $\tau={t}/{n}$ and accordingly $\tau_n^{\star}={t_n^{\star}}/{n}$. 
Based on large-deviation result on the sum of independent and identically distributed (i.i.d.) random variables uniformly distributed on $[0,1]$ (see Appendix \ref{app.ldt}), we establish the asymptotics of $\tau_n^{\star}$ in what follows.

\begin{theorem}\label{thm.convergence}
	For any given $\EEE\in (0,\frac{1}{2})$, the parameter of the optimal shaping region satisfies
	\begin{flalign}\label{eq.tau}
	\tau_n^{\star}=\EEE+\frac{1}{\mu^*}\frac{1}{n}+o\left(\frac{1}{n}\right),
	\end{flalign}
	where $\mu^*$ is the unique positive solution to the following equation
\begin{flalign}\label{eq.solution}
\EEE=\frac{1}{\mu}-\frac{1}{\exp\left(\mu\right)-1}.
\end{flalign}
\end{theorem}	
\begin{IEEEproof}
	See Appendix~\ref{app.convergence}.
\end{IEEEproof}
Theorem \ref{thm.convergence} shows that the optimal shaping parameter $\tau_n^{\star}$ converges to $\alpha$ at a rate of $\frac{1}{\mu^*}\frac{1}{n}$, where $\mu^*$ is also the exponential decay rate of the following probability density function:
\begin{flalign}\label{eq.max_ent}
p_{X}^*(x)=\frac{\mu^*}{1-\exp\left(-\mu^*\right)}\exp\left(-\mu^*x \right),\quad 0\le x\le 1,
\end{flalign}
that maximizes the differential entropy of the input $X$ under the support constraint $X\in [0,1]$ and the average-intensity constraint $\mathbb{E}\left[ X\right]=\alpha$ \cite{Lapidoth2009Capacity}.

As a consequence of Theorem \ref{thm.convergence}, we may approximate the shaping parameter $t_n^{\star}$ by $ n\EEE+{1}/{\mu^*}$, which facilitates us to determine the optimal shaping region in the application of lattice codes, rather than numerical methods.

\subsection{Finite-Blocklength Analysis of Maximum Shaping Gain}
Relying on Theorem \ref{thm.convergence}, a second-order expansion of the maximum shaping gain can be obtained as follows.

\begin{theorem}\label{thm.shaping gain}(Second-Order Asymptotics) For any given $\EEE\in (0,\frac{1}{2})$, the maximum shaping gain $\overline{\mathtt{SG}}_{\textnormal{VLC}}(n;\alpha)$ achieved in the Euclidean-$n$ space satisfies
\begin{flalign}\label{eq:shaping_gain_approximation}
	\log\overline{\mathtt{SG}}_{\textnormal{VLC}}(n;\alpha)=\hh_{\text{max}}(\alpha)-\log(2\EEE)-\frac{\log(n)}{2n} +\frac{\omega_{\alpha}}{n}
+o\left(\frac{1}{n}\right)
\end{flalign}
where $\hh_{\text{max}}(\alpha)$ is the differential entropy of $p_{X}^*(x)$ in Eq. \eqref{eq.max_ent}, and the constant $\omega_{\alpha}$ is given by 
\begin{flalign}
    \omega_{\alpha}=
1-\frac{1}{2}\log\left(2\pi\left( -\mu^*+2\alpha \mu^*+\left(\mu^* \right)^2\alpha\left(1-\alpha\right)\right)
\right).
\end{flalign}
\end{theorem}
\begin{IEEEproof}
	See Appendix~\ref{app.shaping gain}.
\end{IEEEproof}

It is straightforward that we can use the second-order expansion in \eqref{eq:shaping_gain_approximation} to approximate the maximum shaping gain in finite dimensions. To the authors' best knowledge, this is the first quantitative result on measuring how much shaping gain can be obtained for the VLC in the regime of short blocklengths, which may be helpful for system design in the field. It is worth noting that the performance limit of lattice codes in the finite-blocklength regimes can be further clarified by combining with Theorem \ref{thm.shaping gain} and the dispersion result of unconstrained Gaussian channel \cite{Ingber2013Finite}.

%\begin{remark}[Comparison With Dispersion of Unconstrained Gaussian Channel]
%\end{remark}

%\begin{remark}[Duality between Optimal Shaping Region and The]
	%Before presenting the optimal shaping region, it is necessary to clarify the duality between the one-dimensional maximum entropy distribution and the high-dimensional uniform distribution. Since the characteristics of the high-dimensional uniform distribution are determined by the geometric characteristics of the support set, it can also be considered as a geometric statement of the maximum entropy distribution.
%\end{remark}

%\subsection{Convergence of Marginal Distribution}
%
%Next, we investigate the marginal PDF of the random vector uniformly distributed on $\mathscr{T}_n\left(t_n^{\star}\right)$, which is suitable for approximating lattice codes at a relatively high rate~\cite{Forney1989Multidimensional1}. Due to the permutation-invariance property of $\mathscr{T}_n\left(t_n^{\star}\right)$, we denote the marginal PDF as $f_{X_1,n}(x)$ for $x\in [0,1]$. The limiting behavior of $f_{X_1,n}(x)$ is characterized in the following theorem.

Theorem 3 also demonstrates an interesting connection between the maximum-entropy distribution and the optimal shaping region for our considered channel. It can be further proved by Theorem 3 that any marginal probability density function of the random vector uniformly distributed on $\mathscr{T}_n(t^\star_n)$ converges to the maximum-entropy distribution \eqref{eq.max_ent}, which reveals the superiority of the geometric shaping at high SNRs.

\begin{figure}[h]
	\centering
	\begin{overpic}[width=0.45\textwidth]{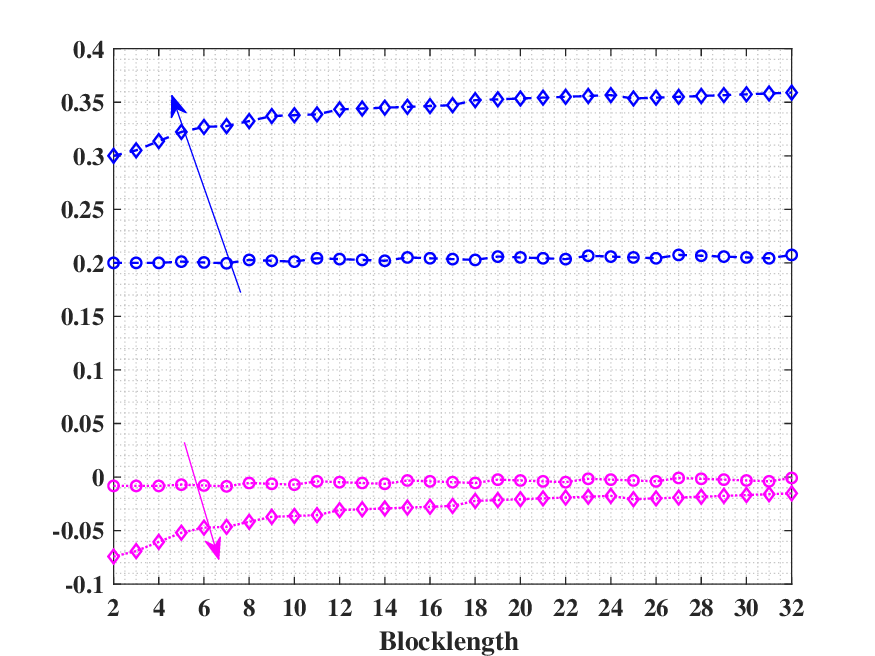}	
		\put(25,51){\small\bfseries\color{black}{$t_n^{\star}-n\EEE,\, \EEE=0.2,\,0.3$}}
		\put(28,11){\small\bfseries\color{black}{$t_n^{\star}-n\EEE-\frac{1}{\mu^*},\, \EEE=0.2,\,0.3$}}
	\end{overpic}
	\centering \caption{The approximation error $t_n^{\star}-n\EEE-{1}/{\mu^*}$ versus various blocklengths $n$ and constraint parameters $\EEE$.}
	\label{Fig.tn}
\end{figure}

We end this subsection by relevant numerical results.
In Fig. \ref{Fig.tn}, we plot the absolute approximation error $t_n^{\star}-n\EEE-{1}/{\mu^*}$ for blocklengths $n$ from $2$ to $32$, which shows a good approximation performance even at very short blocklengths. 

\begin{figure}[!htbp]
	\centering
	\resizebox{8cm}{!}{\includegraphics{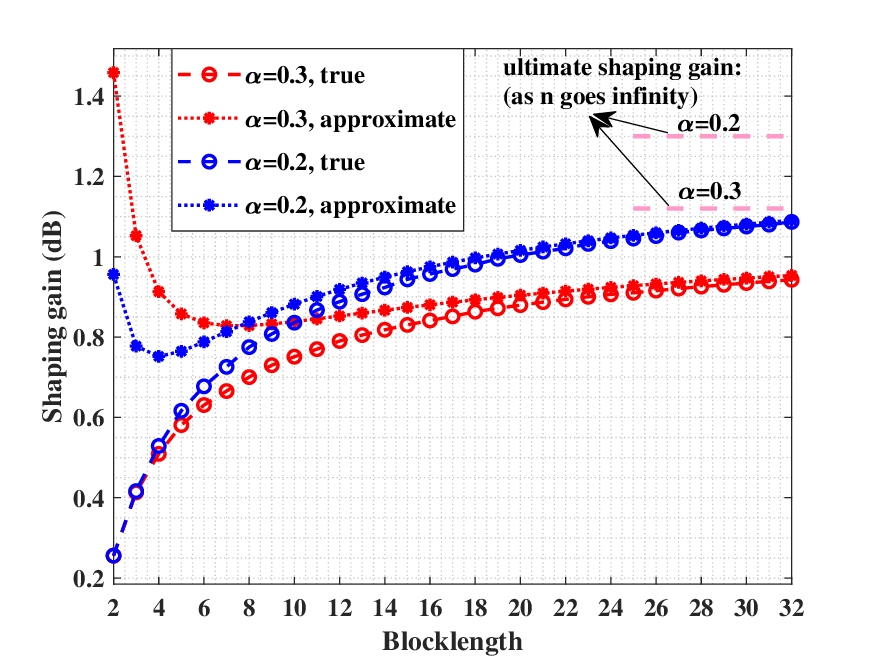}}
	\centering \caption{True shaping gain and the second-order approximation with different constraint parameters.}
	\label{Fig.siso_shaping_gain}
\end{figure}

In Fig. \ref{Fig.siso_shaping_gain}, we plot the true maximum shaping gains and the approximated shaping gains versus various dimensions $n$ and different intensity constraint parameters $\alpha$. The results show that the true maximum shaping gain can be effectively approximated by its second-order expansion. When $n\ge16$, the approximate error is no larger than $0.1$ dB.

\begin{figure}[!htbp]
	\centering
	\resizebox{8cm}{!}{\includegraphics{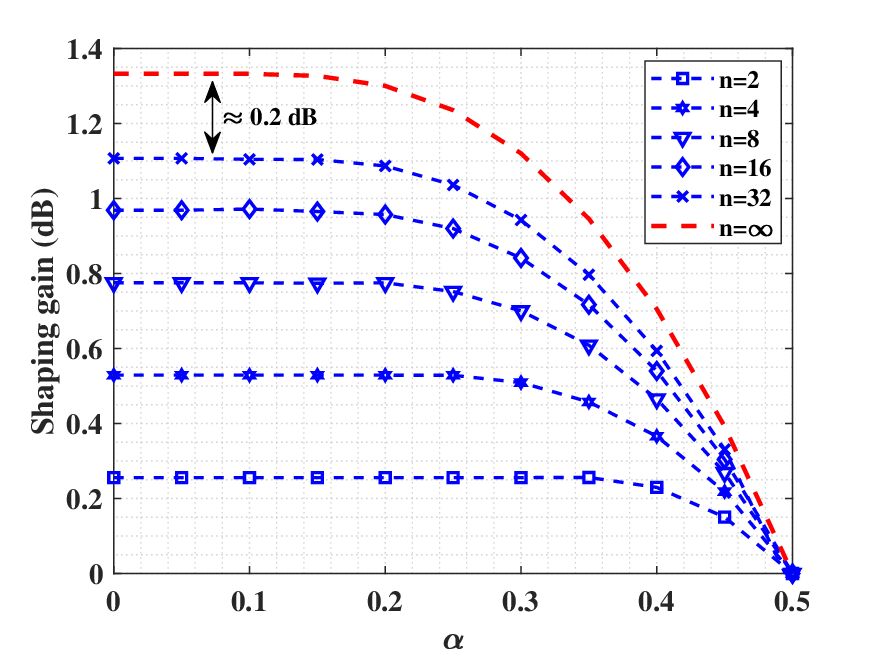}}
	\centering \caption{The relationship between the maximum shaping gain $\overline{\mathtt{SG}}_{\textnormal{VLC}}(n;\alpha)$ and the constraint parameter $\alpha$ with different dimensions.}
	\label{Fig.shaping_gain}
\end{figure}

Fig. \ref{Fig.shaping_gain} provides the relationship between the maximum shaping gain $\overline{\mathtt{SG}}_{\textnormal{VLC}}(n;\alpha)$ and the constraint parameter $\alpha$ in various dimensions. It can be seen from Fig. \ref{Fig.shaping_gain} that for any $\alpha\in\left(0,\frac{1}{2}\right)$, the gap between the ultimate shaping gain and the maximum shaping gain obtained in the $32$-dimensional space is no larger than $0.2$ dB, which indicates that most of the shaping gain can be obtained at low dimensions. 

\subsection{Extension to APP-Limited Quadrature Gaussian Channel}
Based on the above result, we also characterize the second-order asymptotics of the maximum shaping gain for \textit{the average- and peak-power-limited (APP-limited) quadrature Gaussian channel} \cite{Kschischang1994Optimal,Shamai1995capacity}, whose channel output over $\frac{n}{2}$ channel uses is given by
\begin{flalign}
    \mathbf{r}_{\textnormal{E}}=\mathbf{x}_{\textnormal{E}}+\mathbf{z}_{\textnormal{E}},
\end{flalign}
where the $\frac{n}{2}$-dimensional complex-valued transmitted signal $\mathbf{x}_{\textnormal{E}}=\left(\tilde{x}_1,\cdots,\tilde{x}_{\frac{n}{2}} \right) $ 
is equiprobably selected from the constellation $\mathcal{\widetilde{X}} \subseteq \mathbb{C}^{{n}/{2}}$ of size $M$.
Due to the average and peak power constraints, 
the transmitted constellation $\mathcal{\widetilde{X}}$ is required to satisfy
\begin{subequations}\label{eq.constraint2}
	\begin{align}
 	\left| \tilde{x}_k \right|^2 \le 1,& \, \forall k\in\left\lbrace 1,\cdots,\frac{n}{2}\right\rbrace \label{eq.constraint2A}\\ 
	\dfrac{1}{\frac{n}{2}M}\sum_{\mathbf{x}_{\textnormal{E}}\in\mathcal{\widetilde{X}}}\sum_{k=1}^{n/2} \left| \tilde{x}_k \right|^2 \le \const{P}.&\label{eq.constraint2B}
	\end{align}
\end{subequations}

For the APP-limited quadrature Gaussian channel, the shaping gain of a given closed region $\mathscr{R}$ of the corresponding Euclidean $n$-space is defined as
\begin{flalign}
 \mathtt{SG}_{\text{E}}\left( \mathscr{R} \right)= \frac{\PAPR{\mathscr{R}}}{3},
\end{flalign}
where the quantity $3$ is the peak-to-average-power ratio (PAPR) of the $n$-cube \cite{Kschischang1994Optimal}.
For simplicity, the maximum shaping gain within the blocklength $\frac{n}{2}$ of the APP-limited quadrature Gaussian channel is denoted by $\overline{\mathtt{SG}}_{\text{E}}(\frac{n}{2};\const{P})$, which is achieved by the $n$-dimensional real-valued truncated polydisc (see \cite{Kschischang1994Optimal} for a detailed description). Combining \cite[(9)]{Kschischang1994Optimal} and Theorem \ref{thm.shaping gain}, we can readily derive the second-order asymptotics of the maximum shaping gain for the APP-limited quadrature Gaussian channel in the following corollary, which may be of independent interests.

%\begin{flalign}
%\text{PAR2}[\Theta_{2n}(np_n)]=\begin{cases}1/\psi_n(1)=n+1,\quad&0<np_n \leq1\\1/\psi_n(np_n),\quad&np_n>1.\end{cases}
%\end{flalign}
%\begin{flalign}
%\vol\left( \Theta_{2n}(n p_{n})\right) =\left(\pi\right)^n\phi_n(n p_{n}),
%\end{flalign}
%
%\begin{flalign}
%	\gamma_{s}[\Theta_{2n}(np_n)]=\frac{\pi}{6}\times\frac{\phi_{n}(np_n)^{1/n}}{\psi_{n}(np_n)}.
%\end{flalign}

\begin{corollary}\label{cor. RF shaping gain}
The maximum shaping gain of the APP-limited quadrature Gaussian channel satisfies
	\begin{flalign}
& \overline{\mathtt{SG}}_{\textnormal{E}}(\frac{n}{2};\const{P})\nonumber \\
 =&\,\frac{\pi}{3}
 \overline{\mathtt{SG}}_{\textnormal{VLC}}(n;\const{P})\nonumber  \\
 =&\,0.200+\hh_{\text{max}}(\const{P})-\log(2\const{P})-\frac{\log(n)}{2n} 
 \nonumber\\
 &\,+\frac{\omega_{\const{P}}}{n}
+o\left(\frac{1}{n}\right) \quad ( \textnormal{in dB}).
	\end{flalign}
 \end{corollary}

%\begin{IEEEproof}
%	See Appendix~\ref{app.RF shaping gain}.
%\end{IEEEproof}

We especially point out that, by letting $\const{P} \to 0$ and $n\to +\infty$, Corollary 1 coincides with the well-known results that  the ultimate shaping gain (with respect to OSNR) for the VLC channel with only an average-intensity constraint is $1.33$ dB (see \cite{Shiu1999Shaping}), while that (with respect to electrical SNR) for the AWGN channel with only an average-power constraint is $1.53$ dB.
%\rev{RF shaping gain}

%\gjn{arg1}

%In the following, such the connection will be further clarified. Define $\mathbf{X}=\left( X_1, \cdots, X_n\right)$ as a random vector uniformly distributed on $\mathscr{T}_n\left(t_n^{\star}\right)$. 
%Due to the permutation-invariance property of $\mathscr{T}_n\left(t_n^{\star}\right)$, we can investigate the marginal probability density function of the random variable $X_1$, which is denoted by $f_{X_1,n}(x)$ for $x\in [0,1]$. A convergence property of the marginal distribution of the optimal shaping region is given in Proposition \ref{pro.marginal}. 
%%The limiting behavior of $f_{X_1,n}(x)$ is characterized in the following theorem.
%\begin{proposition}\label{pro.marginal} 
%	\textit{As $n\to +\infty$, any marginal distribution of the random vector uniformly distributed on $\mathscr{T}_n\left(t_n^{\star}\right)$ follows the maximum-entropy distribution \eqref{eq.max_ent}, i.e.,}
%	\begin{flalign}
%	f_{\infty}(x)=\lim_{n\to \infty}f_{X_1,n}(x)=p_{X}^*(x),
%	\end{flalign}
%\end{proposition}
%\begin{IEEEproof}
%See Appendix~\ref{app.marginal}.
%\end{IEEEproof}
%Remark~\ref{pro.marginal} reveals that any marginal of the optimal shaping region will behave like the maximum-entropy distribution as the blocklength tends to infinity.

\section{Optimally-Shaped Constellation Based on Construction B Lattices}\label{Sec.coding}
In Section \ref{Sec.shaping}, we derive the second-order asymptotics of the optimal shaping parameter $t_n^{\star}$ and the maximum shaping gain $\overline{\mathtt{SG}}_{\textnormal{VLC}}(n;\alpha)$ for the VLC channel under the dual intensity constraints, which answers the question that how much shaping gain can be achieved by finite-dimensional geometric shaping. 
In order to improve the coding gain, one approach is to employ an inequiprobable ASK constellation in conjunction with powerful channel codes like LDPC codes. However, it may not be suitable for indoor VLC scenarios due to the need of distribution matching and time-consumed iterative decoding algorithms.

%\rev{Main contribution of this section}
Therefore, in this section, we propose a general signaling framework by constructing geometrically-shaped constellations, which has flexible normalized rates, low implementation complexity, and a substantial OSNR gain as compared with existing schemes.

%we propose a constellation design via using points from a densely-packing lattice from shaping gains brought by geometrically shaping regions and coding gains
%Therefore, in this section,
%via using points from a densely-packing lattice from shaping gains brought by geometrically shaping regions and coding gains by s, XX 
%we propose a general framework, which has flexible normalized rates and a substantial OSNR gain as compared with existing schemes. It could be realized at the cost of relatively low implementation complexity.}

%\rev{However, there is no simple expression and a systematic understanding of the optimal shaping parameter, which is numerically computed in \cite{Chen2020FFT}.}

\subsection{Basic Idea of Constructing Geometrically-Shaped Constellations}
\subsubsection{Denser Packing via Construction B Lattice}
%In our design, we exploit the good algebraic structure of the \gjn{Construction B lattice $H_{n}$} to augment the coding gain.

It is noted that the most of the ultimate shaping gain $\overline{\mathtt{SG}}_{\textnormal{VLC}}(+\infty;\alpha)$ can be achieved in finite dimensions (as illustrated in Fig. \ref{Fig.shaping_gain}), while there is still much room for further improvement of the coding gain. In \cite{Chen2020FFT}, both the shaping and the coding of the truncated cubic constellation (TCC) are carried out by the checkerboard lattice $D_n$, of which a nominal coding gain of $1.5$ dB (with respect to the OSNR definition \eqref{eq:snr}) is attained as compared with the ASK constellation.

To address the limitation of the $D_n$ lattice on the coding gain of the TCC, we leverage the densely-packing structure of the Construction B lattice, or more generally, the union of cosets of the Construction B lattice 
	\begin{flalign}
		\Lambda_{n}=\bigcup_{\mathbf{a}\in \mathcal{A}} \left( 2H_{n} + \mathbf{a}\right),
		\label{eq.construction B coset}
	\end{flalign}
	where $\mathcal{A}$ is the union of coset representatives. It has been found that many dense lattice packings in low dimensions can be constructed by translating a Construction B lattice, such as $E_8$ lattice, $BW_{16}$ lattice, and Leech lattice \cite{Forney1989bounded}. The good choice of $\mathcal{A}$ varies greatly with the blocklength $n$, and hence, is not specified here.

%and fortunately, a bounded distance decoding algorithm proposed in \cite{Forney1989bounded} can assist us in demodulation. Moreover, we employ a bounded distance decoding algorithm proposed in \cite{Forney1989bounded} to enhance our demodulation process.
%\rev{Fortunately, a bounded-distance decoding algorithm proposed in \cite{Forney1989bounded} to simplify demodulation.}

\subsubsection{Coarsely Shaping and Finely Coding}
To integrate the coding gain of the finitely-shifted Construction B lattice and the nearly-maximum shaping gain in the constellation construction, a natural way is selecting $M$ points from the lattice $\Lambda_n$ within the optimal shaping region $ \mathscr{T}_n \left(t_n^{\star}\right)$ (scaling by an appropriate factor). However, directly enumerating $M$ points from the intersection of a delicately-constructed lattice $\Lambda_n$ with a nontrivial Voronoi cell and the optimal shaping region is challenging, and a look-up table may lead to an exponential growth of the computation complexity and the storage complexity as the normalized rate $\beta$ increases.

To address this issue, resorting to the idea of \textit{coarsely shaping and finely coding} and the structure of Construction B lattices, we construct the multidimensional constellation with a near-optimal shape by assigning the task of constellation shaping to a sublattice of the embedded Construction B lattice.

% , called \textit{optimally-shaped \gjn{finitely-shifted Construction-B} constellation}, based on the aforementioned lattice construction in \gjn{Eq. \eqref{eq.construction B}} and the idea of , where the constellation shaping function is assigned to a 

\begin{figure}[h]
	\centering
	\resizebox{8cm}{!}{\includegraphics[width=10cm]{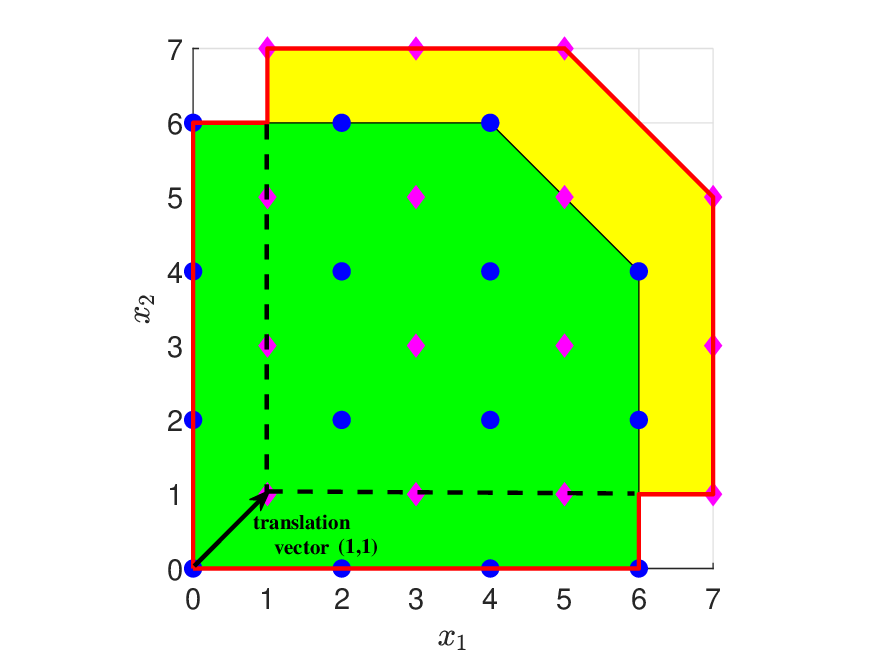}}\\
	\caption{Coarsely shaping and finely coding for selecting $D_2$ points from a truncated cube.}
	\label{Fig.coarsely shaping}
\end{figure}

For better readability, here we briefly illustrate the basic idea of coarsely shaping and finely coding by a two-dimensional example in Fig. \ref{Fig.coarsely shaping}. Suppose we select $M$ points from the $D_2$ lattice, whose boundary is required to look like a given truncated cube. Note that  $D_2$ can be represented in a Construction A form
\begin{flalign}
D_{2}=2\mathbb{Z}^{2}+\mathcal{C}_{2},
\label{eq.D2}
\end{flalign}
where the set $\mathcal{C}_{2}=\left\{\bm{0}_2,\bm{1}_2\right\}$. Then a simple construction method is first selecting $M/2$ $2\mathbb{Z}^2$ points from the desired truncated cube (see blue small circles in Fig. \ref{Fig.coarsely shaping}) and then translating those points by the vector $\bm{1}_2\in \mathcal{C}_{2}$ (see magenta small diamonds in Fig. \ref{Fig.coarsely shaping}). It can be seen that the overall boundary almost maintains the desired shape, say coarsely shaping, while simultaneously all points are selected from $D_2$ lattice which determines the MED of the constructed constellation, say finely coding.

\subsection{Framework of Constellation Construction}\label{sec:constellation_construction}
%\subsubsection{Constellation Construction}
%\rev{Motivated by the above idea and the Eqs. \eqref{eq.construction B} and \eqref{eq.construction B coset}, the boundary of the high-dimensional constellation $\mathcal{X}$ to be designed can be coarsely controlled by restricting the component $D_{n}$ lattice points in the desired shaping region $\mathscr{T}_{n}(t^{\star})$. Based on Theorem 2, we may replace the optimal parameter $t^{\star}$ by $n\alpha+1/\mu^{\ast}$. In the following, the construction of the proposed constellations} are presented in details. 

Combining the construction \eqref{eq.construction B} of $H_n$, and analogue to the above two-dimensional example, our constellation construction is formally given in what follows:
\begin{flalign}
    \mathcal{L}_n=& \bigcup_{\mathbf{a}\in \mathcal{A}}
    \underbrace{4 \mathcal{D}_{\text{shape}}}_{\text{shaping part}} + \underbrace{2 \mathcal{C} + \mathbf{a}}_{\text{coding part}}, \label{eq.constellation framework}
\end{flalign}
where the shaping set $\mathcal{D}_{\text{shape}}$ is chosen from the $D_n$ lattice and used to control the overall shape of $ \mathcal{L}_n$.
It is clear that the so-constructed constellation $\mathcal{L}_n$ is obtained via translating (at a small scale) $\mathcal{D}_{\text{shape}}$ by the quotient set $\left[\Lambda_{n}/4D_{n}\right]$. In the following, three crucial issues in constructing $\mathcal{L}_n$ are clarified in detail.
%\begin{flalign}
%	\mathcal{L}_{n}=\bigcup_{\mathbf{a}\in A} \left( r\mathcal{H}_{n}+\mathbf{a} \right),
%	\label{eq.construction B constellation}
%\end{flalign}
%where the set \gjn{$\mathcal{H}_{n}\triangleq 2\mathcal{D}_{n}\left({H}^*,2{L}^*,M_s\right)+C$}.
%Recall 
%$$
%H_n= 2D_n +\mathcal{C}
%$$
%shaping set  $\mathcal{D}_{\text{shape}}(\beta_s,\alpha) \subseteq D_n$. 

\subsubsection{Choice of $\mathcal{D}_{\text{shape}}$}\label{sec:shaping_set}
Assume that there are $2^{k_a}$ elements in the set $\mathcal{A}$. Let the constants $k_{s}=n\beta-k_c-k_a$ and $M_s=2^{k_{s}}$. Therefore, to get a near-maximum shaping gain, the shaping set $\mathcal{D}_{\text{shape}}$ is required to contain $M_s$ $D_n$ lattice points within the optimal shaping region $\mathscr{T}_n \left(t_n^{\star}\right)$ (with appropriate scaling),
which is exactly the set $\mathcal{D}_{n}\left({H^{\ast}},2{L}^{\ast},M_s\right)$, where the involved parameters $H^{\ast}$ and ${L}^{\ast}$ can be uniquely determined by using Algorithm 1 in \cite{Chen2020FFT} for constructing TCCs.

%The $D_n$-based truncated cube as well as this fast mapping algorithm will also play an important role in our constellation mapping scheme. 

%\subsection{Optimally-Shaped Constellation Based on Construction-B Lattice}
\subsubsection{Fast Constellation Mapping}
It is recognized that a constellation addressing algorithm that quickly maps the message onto the desired constellation point $\bm{\lambda}\in \mathcal{L}_n$ is needed to reduce the implementation complexity of geometric shaping. According to the form of the construction \eqref{eq.constellation framework}, we shall divide the message into three parts, which are mapped onto the shape set $\mathcal{D}_{\text{shape}}$, the component $\left(n,k_c,8 \right)$
code $\mathcal{C}$, and the set $\mathcal{A}$ of coset representatives, respectively. The last two mappings are straightforward and very fast, while fortunately, the mapping onto $\mathcal{D}_{\text{shape}}$ can be efficiently solved by the FFT-assisted decomposable shell mapping (FDSM) in \cite{Chen2020FFT} at the computational cost approximately linear with the blocklength.

For clarity, we summarize the detailed procedures that map a binary sequence $\mathbf{b}$ onto the constellation point $\bm{\lambda}\in \mathcal{L}_{n}$ in the following.
\begin{itemize}
	%	\item \textit{Step 1: Calculating shaping parameter.} The shaping code consists of the checkerboard lattice points in a truncated cube
	%	Calculate the shaping parameter $t_{24}^*=24\EEE-{1}/{\mu}^*$, which is provided in Theorem \ref{thm.convergence} and Eq. \eqref{eq.tau} to determine the optimal shaping region.
	\item \textit{Step 1: Data partition.} 
	Let $\mathbf{b}\in\mathbb{F}_2^k$ be a length-$k$ binary sequence, where $k=n\beta$. Then divide $\mathbf{b}$ into three parts $\mathbf{b}_{s}=\left(b_1,\cdots,b_{k_s}\right)$, $\mathbf{b}_{c}=\left(b_{k_s+1},\cdots,b_{k_s+k_c}\right)$, and $\mathbf{b}_{a}=\left(b_{k_s+k_c+1},\cdots,b_{k}\right)$.
	\item \textit{Step 2: Coarsely shaping.}
	Map the length-$k_s$ binary sequence  $\mathbf{b}_{s}$ onto some $n$-dimensional lattice point $\mathbf{d}=\left( d_1,\cdots,d_{n}\right)\in \mathcal{D}_{n}\left({H^*},2{L^*},M_s\right)$ by the FDSM algorithm \cite[Algorithm 2]{Chen2020FFT}. 
	%\footnote{\gjn{The essence of the algorithm is to create an index of constellation points that share the same power, and then select one point from that indexed set for transmission.}}.
	\item \textit{Step 3: Block coding.}
	Map the length-$k_c$ binary sequence $\mathbf{b}_{c}$ onto some codeword $\mathbf{c}=\left(c_1,\cdots,c_{n}\right) \in \mathcal{C}$.
	\item \textit{Step 4: Coset mapping.}
	Map the length-$k_a$ binary sequence $\mathbf{b}_a$ onto some coset representative $\mathbf{a}\in\mathcal{A}$ via an enumeration approach. Then the desired constellation point is obtained as $\bm{\lambda}=4\mathbf{d}+2\mathbf{c}+\mathbf{a}$.
	%If $b_k=0$, then $\bm{\lambda}=4\mathbf{d}+2\mathbf{g}$. Otherwise, $\bm{\lambda}=4\mathbf{d}+2\mathbf{g}+\tilde{\bm{\xi}}$, where $\tilde{\bm{\xi}}$ is determined by Eq. \eqref{eq.xi_optical}.
\end{itemize}

\subsubsection{Constellation Scaling}
For simplicity, in the above we construct the desired constellation based an unnormalized Construction B lattice, which should be scaled to meet the input constraints \eqref{eq.constraint}. Here we will briefly discuss how to evaluate the scaling factor $\kappa\in \mathbb{R}_+$.

Note that the peak value of the proposed constellation $\mathcal{L}_{n}$ is given by
\begin{flalign}
	&\max_{\bm{\lambda}\in\mathcal{L}_{n}}\left\|\bm{\lambda}\right\|_{\infty}\\
	=&\max_{\mathbf{d}\in\mathcal{D}_{n}\left({H}^*,2{L}^*,M_s\right)
		\atop \mathbf{c} \in \mathcal{C},\, \mathbf{a} \in \mathcal{A}}\left\|4\mathbf{d}+2\mathbf{c}+\mathbf{a}\right\|_{\infty}\nonumber\\
	=&4H^*+2+\max_{\mathbf{a}\in\mathcal{A}}\left\|  \mathbf{a}\right\|_{\infty}.\label{eq.peak power2}
\end{flalign}
Similarly, the average intensity of $\mathcal{L}_{n}$ is given by
\begin{flalign}
&\frac{1}{nM}\sum_{\mathbf{x}\in\mathcal{L}_{n}} \left\|\mathbf{x}\right\|_{1}\nonumber \\
= &\frac{1}{nM}\sum_{\mathbf{d}\in\mathcal{D}_{24}\left({H}^*,2{L}^*,M_s\right) \atop \mathbf{c} \in \mathcal{C}, \, \mathbf{a} \in \mathcal{A}} \left\|4\mathbf{d}+2\mathbf{c}+\mathbf{a}\right\|_{1} \\
=& \frac{4}{M_s}\sum_{\mathbf{d}\in\mathcal{D}_{n}\left({H}^*,2{L}^*,M_s\right)} \left\|\mathbf{d}\right\|_{1}+1+\dfrac{\sum_{\mathbf{a} \in \mathcal{A}}\left\|\mathbf{a}\right\|_{1}}{2^{k_a}},\label{eq.average power}
\end{flalign}
where Eq. \eqref{eq.average power} follows from the fact that the uniformity of any linear code.

Thus, to satisfy the intensity constraints \eqref{eq.constraint}, the scaling factor $\kappa$ should be set as
	\begin{equation}\label{eq.scaling factor}
	\kappa=1/ \max {\left\lbrace \max_{\bm{\lambda}\in\mathcal{L}_{n}}\left\|\bm{\lambda}\right\|_{\infty},  \frac{1}{nM}\sum_{\bm{\lambda}\in\mathcal{L}_{n}} \left\|\bm{\lambda}\right\|_{1}/ \EEE \right\rbrace}.
	\end{equation}

 We would like to point that, for simplicity of implementation, the maximum coordinate $H^{\ast}$ and the average $\ell_1$-norm of $\mathcal{D}_{n}\left({H}^*,2{L}^*,M_s\right)$ can be roughly estimated by using the optimal shaping region $\mathscr{T}_n \left(t_n^{\star}\right)$ via continuous approximation (or say, the Minkowski-Hlawka theorem \cite[Thm. 1]{Loeliger1997Averaging}).

\subsection{Demodulation Algorithm}\label{Subsubsec.demodulation}
With perfect CSI at the receiver, the maximum likelihood (ML) demodulator performs the minimum-distance detection as
\begin{equation}\label{eq.ML demldulation}
	\hat{\bm{\lambda}}=\argmin_{\bm{\lambda}\in \mathcal{L}_{n}}
	\parallel\bm{\lambda}-\mathbf{r}/\kappa\parallel_2.
	%=\argmin_{\mathbf{x}\in \mathcal{L}_{24}}
	%\parallel\mathbf{x}-\mathbf{y}\parallel_2,
\end{equation}
For simplicity, we let $\mathbf{y}=\mathbf{r}/\kappa$. However, for the used constellation $\mathcal{L}_n$ with a finite cardinality, an irregular boundary, and a complicated multi-layer construction in Eq. \eqref{eq.constellation framework}, the ML demodulation is not feasible in practice due to the complexity limitation. For this reason, we resort to a bounded-distance decoding algorithm for Construction B lattices, which was proposed in \cite{Forney1989bounded} and shown to have near-optimal performance with respect to the symbol error probability.
%For instance, when the bit rate is $\beta=5$bpcu, the cardinality of the constellation is $2^{24\times 5}=2^{120}$. It obviously cannot be decoded by traversal search.  
%For the Leech lattice decoding, many decoding algorithm has been proposed. Vardy and Be'ery provide ML decoding of $4\times Q_{24}$ using ML hexadecoder \cite{Vardy1993Maximum}. Forney proposed a bounded distance decoding algorithm of $H_{24}$. The algorithm has the same error performance as the ML decoding when the distance between the received signal and the transmitted signal is less than $\sqrt{2}$ \cite{Forney1989bounded}. A bounded distance decoding algorithm of $Q_{24}$ is proposed by Amrani and Be'ery \cite{Amrani1994Efficient}. 
\subsubsection{Bounded-Distance Decoding}

For any given vector $\mathbf{y}\in \mathbb{R}^n$ and a Construction B lattice $H_{n}$, the bounded-distance decoding algorithm \cite[Algorithm 2]{Forney1989bounded} outputs the  point $\mathbf{h}$ from $H_{n}$ (i.e., infinite constellation) that is closest to $\mathbf{y}$ if $\left\|\mathbf{h}-\mathbf{y}  \right\|\le d_{\min}\left(H_n\right)/2=\sqrt{2}$, i.e., performs
as optimal algorithms within a ball of certain radius. For self-containment, we shall review the main idea of this algorithm in what follows.

It is clear that
\begin{equation}\label{eq.En}
H_n \subseteq	U_{n}=2\mathbb{Z}^{n}+\mathcal{C},
\end{equation}
where $U_{n}$ is a Construction A lattice which is obtained by replacing the single-parity code in constructing $H_n$ to $\left\{ 0,1\right\}^n$, where the quotient set $\left[U_{n}/H_{n} \right]=\left\{\bm{0}_{n},\left( \bm{0}_{m},2,\bm{0}_{n-m-1}\right)\right\}$ with an arbitrary integer $m$ ranging from $0$ to $n-1 $. The aforementioned algorithm first searches the closest point from $U_n$, which can be quickly solved by combining integer forcing and soft decoding of the binary linear block code $\mathcal{C}$ due to the property of Construction A lattices \cite{conwaysloane99_1}. Then the algorithm translates such the closest point from $U_n$ in some way to ensure that the final output point is selected from the sublattice $H_n$. 

%
%Let us consider the ML decoding problem for the \gjn{Construction B} lattice  \gjn{, which is the same as a construction A expression by replacing }  We first review the following construction-A expression of \gjn{$E_{n}$} lattice
%
%%with the $\left( 24,12,8\right) $ Golay code as the binary linear code \cite{conwaysloane99_1}.
%Note that \gjn{$D_{n}$} is a sublattice of \gjn{$\mathbb{Z}^{n}$} that consists of all the points with even sum. By comparing Eqs. \gjn{\eqref{eq.construction B}} and \eqref{eq.En}, it is easily seen that \gjn{$H_{n}$} is a sublattice of \gjn{$E_{n}$}, and the corresponding coset \gjn{$\left[E_{n}/H_{n} \right]=\left\{\bm{0}_{n},\left( \bm{0}_{m},2,\bm{0}_{n-m-1}\right)\right\}$}, where $m$ can be chosen to be any integer number between $0$ and \gjn{$\left( n-1\right) $}. 

%If the noise vector is bounded, then the decoding problem for \gjn{$H_{n}$} with a \gjn{Construction B} structure can be converted into that for \gjn{$E_{n}$} with a \gjn{Construction A} structure, for which the closest point search is equivalent to soft decoding of the \gjn{binary block code $\mathcal{C}$} \cite{conwaysloane99_1}.

\subsubsection{Demodulation Procedure}
Motivated by the above fact, the demodulation of our proposed constellation $\mathcal{L}_n$ can be implemented via a lattice decoding approach.  The detailed procedures are summarized in the following.

\begin{itemize}
	\item \textit{Step 1: Closest point search in $U_{n}$} \cite{conwaysloane99_1}. 
	For any $\mathbf{a}\in\mathcal{A}$, let $\mathbf{w}(\mathbf{a})=(\mathbf{y}-\mathbf{a})/2$. Let $\hat{\mathbf{u}}(\mathbf{a})=2\hat{\mathbf{z}}+\hat{\mathbf{c}}$ be the closest point in $U_{n}$ to $\mathbf{w}(\mathbf{a})$, where the integer component $\hat{\mathbf{z}}$ and the binary linear block codeword $\hat{\mathbf{c}}$ are computed by using the decoding algorithm for Construction A lattices \cite[p. 450]{conwaysloane99_1}, along with the soft decoding algorithm for the binary linear block code $\mathcal{C}$.
	%The estimated binary data $\mathbf{\hat{b}}_g=\left(\hat{b}_{k_1+1},\cdots,\hat{b}_{k_1+k_2} \right)$ is also obtained after inputing $\mathbf{\hat{g}}$ to the Golay decoder.
	%Scale the received signal by $\mathbf{y}=\mathbf{r}/\kappa$. Find the closest point $\mathbf{e}$ for $\mathbf{y}$ in $E_{24}$ by Algorithm \ref{alg.E24 decoding}, the estimated Golay code $\mathbf{\hat{g}}$, and the Golay decoding output $\mathbf{\hat{b}}_g=\left(\hat{b}_{k_1+1},\cdots,\hat{b}_{k_1+k_2} \right)$ can also be obtained.
	\item \textit{Step 2: Bounded-distance decoding in $H_{n}$} \cite{Forney1989bounded}. Let $\hat{\mathbf{h}}(\mathbf{a})=\hat{\mathbf{u}}(\mathbf{a})$. If the integer component $\hat{\mathbf{z}}$ has an even sum, we directly output $\hat{\mathbf{h}}(\mathbf{a})$ as the bounded-distance decoding result for $H_n$. Otherwise, search the smallest index $i$ satisfying
	\begin{equation}\label{eq.coordinate index}
		i=\argmax_{j\in \left\lbrace 1,2,\cdots,n \right\rbrace} \mid w_j(\mathbf{a})-\hat{u}_j(\mathbf{a}) \mid,
	\end{equation}
	translate $\hat{\mathbf{h}}({\mathbf{a}})$ by
	\begin{equation}
		\hat{h}_i(\mathbf{a})=\begin{cases}
			\hat{u}_i(\mathbf{a})+ 2,&~\text{if }w_i(\mathbf{a})\ge \hat{u}_i(\mathbf{a});\\
			\hat{u}_i(\mathbf{a})- 2,&~\text{if }w_i(\mathbf{a})< \hat{u}_i(\mathbf{a}).\\
		\end{cases}
	\end{equation}
	and then output the modified $\hat{\mathbf{h}}({\mathbf{a}})$.
	
	%\item \textit{Step 3: Coarsely shaping lattice demapping:} Let \gjn{the coarsely shaping lattice point} $\mathbf{\hat{d}}=\left( \mathbf{\hat{h}}-\mathbf{\hat{g}}\right)/2$. Then we map $\mathbf{\hat{d}}$ onto the binary sequence $\mathbf{\hat{b}}_s= \left(\hat{b_1},\cdots,\hat{b}_{k_1}\right)$ by the FFT-assisted decomposable shell demapping (FDSD) algorithm in \cite[Algorithm 3]{Chen2020FFT}.
	\item \textit{Step 3: Bounded-distance decoding in $\Lambda_{n}$ \cite{Forney1989bounded}.} For each $\mathbf{a} \in\mathcal{A}$, compute Euclidean distances between $2\hat{\mathbf{h}}(\mathbf{a})+\mathbf{a}$ and $\mathbf{y}$, and then select the closest one to $\mathbf{y}$ as the demodulator output $\hat{\bm{\lambda}}$.
	%\item \textit{Step 5: Coset demapping.} Estimate the bit $\hat{b}_{k}$ by the coset representatives. If $\hat{\bm{\lambda}}=2\mathbf{\hat{h}}$, $\hat{b}_{k}=0$, while if $\hat{\bm{\lambda}}=2\mathbf{\hat{h}}+\bm{\xi}$, $\hat{b}_{k}=1$. Then the estimated binary data $\mathbf{\hat{b}}=\left(\hat{b}_1, \cdots, \hat{b}_k\right) $ can be obtained.
\end{itemize}

Note that the constellation demapping of the demodulator output $\hat{\bm{\lambda}}$ onto the binary sequence is a simple inverse transform of the constellation mapping, and therefore omitted.

\section{Design Examples and Numerical results} \label{Sec.simulation} 

In this section, we present an illustrative example of our proposed constellation design framework, and then numerically verify its superiority as compared with the existing schemes.
%, and quantify the difference between the capacity and the proposed OSLC scheme to evaluate the error performance.

%\subsection{Optimal Shaping Parameter and Maximum Shaping Gain}

%\subsection{Error Performance}
\subsection{Optimally-Shaped Leech Constellation}
In this subsection, we show a concrete implementation of our proposed constellation design, by which an energy-efficient $24$-dimensional constellation for our considered channel is designed. Note that the densest packing in $24$-dimensional Euclidean space has been proved to be the Leech lattice $\Lambda_{24}$ \cite{Cohn2017sphere}, by which an extra nominal coding gain of $3$ dB (effective coding gain of $2$ dB) may be attained. A method of constructing the Leech lattice via Construction B is introduced in the following.

\subsubsection{Leech Lattice via Construction B}
Denote the Leech half-lattice by $H_{24}$, which can be constructed via Construction B as follows:
\begin{flalign}\label{eq:H24}
	H_{24}\triangleq 4\mathbb{Z}^{24}+2\left(24,23\right)+G_{24}=2D_{24}+G_{24},
\end{flalign}
where the notation $\left(24,23\right)$ refers to the $24$-dimensional parity check codes, and $G_{24}$ is the $\left(24,12,8\right)$ Golay code.
In \cite{conwaysloane99_1, Forney1988Coset, Amrani1994Leech}, a method of constructing the Leech lattice $\Lambda_{24}$ is presented by translating $H_{24}$ as
\begin{flalign}
	\Lambda_{24}=2H_{24}\cup \left( 2H_{24}+\bm{\xi}\right),
	\label{eq.Leech construction}
\end{flalign}
where the $24$-dimensional translation vector $\bm{\xi}=\left(-3,\bm{1}_{23} \right)\in\mathbb{R}^{24}$.

%\gjn{In order to further elucidate the optimally-shaped constellation based on the finitely-shifted Construction B lattice, }

\subsubsection{Construction Procedures}
In line with Section \ref{sec:constellation_construction}, an optimally-shaped Leech constellation (OSLC) can be constructed in what follows.

\begin{figure*}[h]
	\centering
	\resizebox{15cm}{!}{\includegraphics[width=10cm]{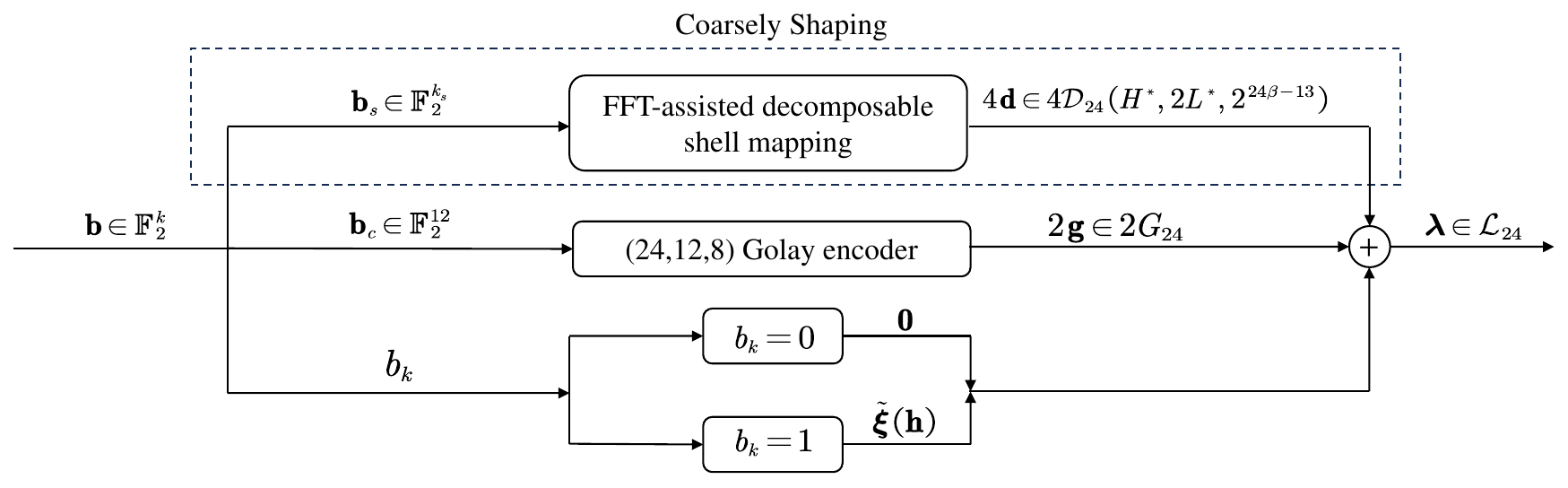}}\\
	\caption{Constellation mapping for the OSLC.}
	\label{Fig.OSLC mapping}
\end{figure*}

Combining Eqs. \eqref{eq.constellation framework} and \eqref{eq:H24}, we define 
\begin{flalign}
	\mathcal{H}_{24}=2\mathcal{D}_{\text{shape}}+G_{24},
\end{flalign}
where the shaping set $\mathcal{D}_{\text{shape}} = \mathcal{D}_{24}\left({H^{\ast}},2{L}^{\ast},2^{24\beta-13}\right) \subseteq D_{24}$ is obtained as described in Section \ref{sec:shaping_set}. To ensure the nonnegativity of the constructed constellation, a slight modification to the next translation operation is needed. Concretely speaking, we construct the OSLC as
	\begin{flalign}
		\mathcal{L}_{24}=\bigcup_{\mathbf{h}\in\mathcal{H}_{24}} \bigcup_{\mathbf{b}\in\left\{ \bm{0}_{24},\tilde{\bm{\xi}}(\mathbf{h}) \right\}} 2\mathbf{h}+\mathbf{b}
 ,
		\label{eq.OSLC}
\end{flalign}
where the modified translation vector $\tilde{\bm{\xi}}(\mathbf{h})$ is given by
\begin{equation}\label{eq.xi_optical}
	\tilde{\bm{\xi}}(\mathbf{h})=
	\begin{cases}
		\left(5,\bm{1}_{23} \right)  &,~2h_1 \text{ mod } 8<3,\\
		\left(-3,\bm{1}_{23} \right) &,~2h_1 \text{ mod } 8>3,\\
	\end{cases}
\end{equation}
and the quantity $h_1$ denotes the first coordinate of $\mathbf{h}\in\mathcal{H}_{24}$. It can be easily verified that, after the above modified translation, so-constructed constellation still lies in the nonnegative orthant and has a cardinality of $M$. For a clearer illustration, we plot procedures of constellation mapping of the OSLC $\mathcal{L}_{24}$ in Fig. \ref{Fig.OSLC mapping}.

%the maximum element of the point in $\mathcal{D}_{24}\left({H}^*,2{L}^*,M_s\right)$ is $H^*$.

 %\gjn{At the receiver, the received signal can be decoded as shown in Algorithm \ref{alg.Leech decoding} according to the procedures in Section \ref{Subsubsec.demodulation}.}
\subsection{Error Performance of $\mathcal{L}_{24}$}
In this subsection, to verify the advantages of our proposed OSLC, we present numerical results on its symbol error rates (SERs). As benchmark schemes, the cubic constellation (i.e., the $24$-times Cartesian product of an ASK constellation) and the $D_{24}$-based TCC proposed in \cite{Chen2020FFT} are considered as well.

%has improved the error performance of the SISO-VLC system under a peak- and an average-intensity constraints \cite{Chen2020FFT}. 
%From the contents of \cite{Chen2020FFT}, the $D_n$-based truncated cubic constellation (TCC) can achieve most of the coding gain when the dimension is $16$. 

\begin{figure*}[!htp]
	\centering
	\subfigure[The SER curves of the OSLC scheme and the benchmarks with $\EEE=0.2$ when $\beta=2$ bpcu and $3$ bpcu.]
	{\label{Fig.r2_beta23_SER}\includegraphics[width=0.45\textwidth]{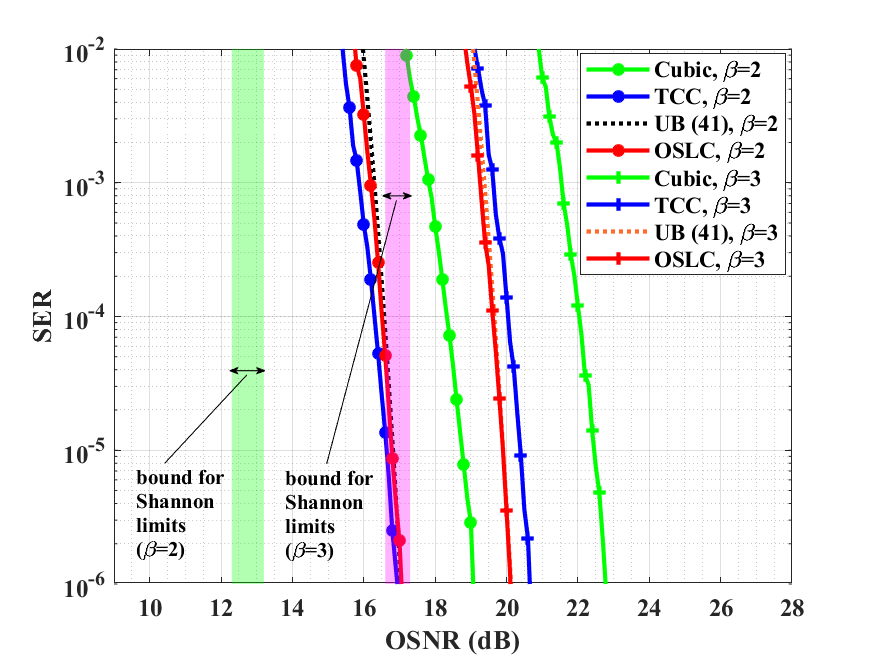}}
	\subfigure[The SER curves of the OSLC scheme and the benchmarks with $\EEE=0.2$ when $\beta=4$ bpcu and $5$ bpcu.]
	{\label{Fig.r2_beta45_SER}\includegraphics[width=0.45\textwidth]{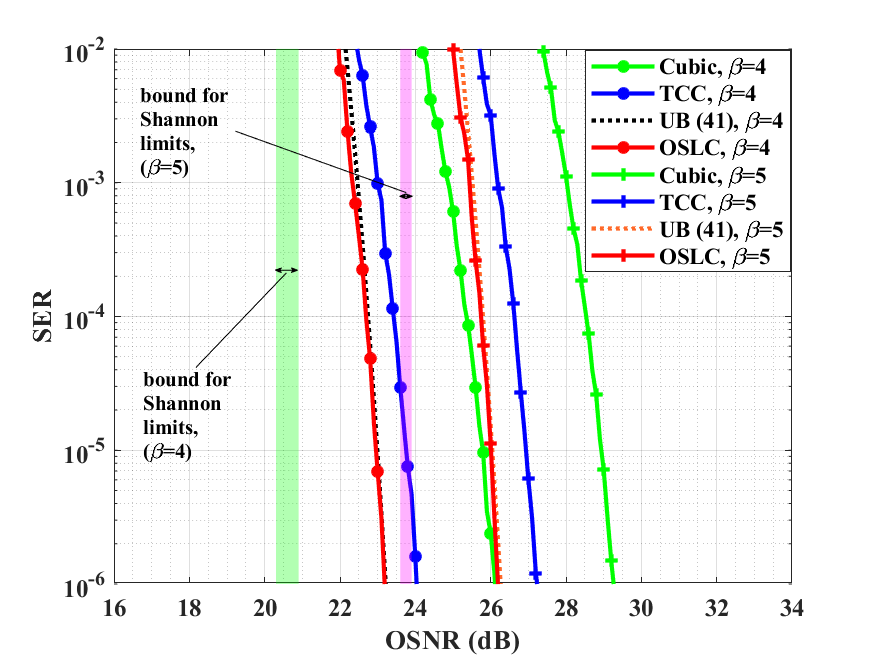}}
	\caption{The SER curves of the OSLC scheme and the benchmarks with $\EEE=0.2$.}
	\label{Fig.r2SER}
\end{figure*}

\begin{figure*}[!htp]
	\centering
	\subfigure[The SER curves of the OSLC scheme and the benchmarks with $\EEE=0.3$ when $\beta=2$ bpcu and $3$ bpcu.]
	{\label{Fig.r3_beta23_SER}\includegraphics[width=0.45\textwidth]{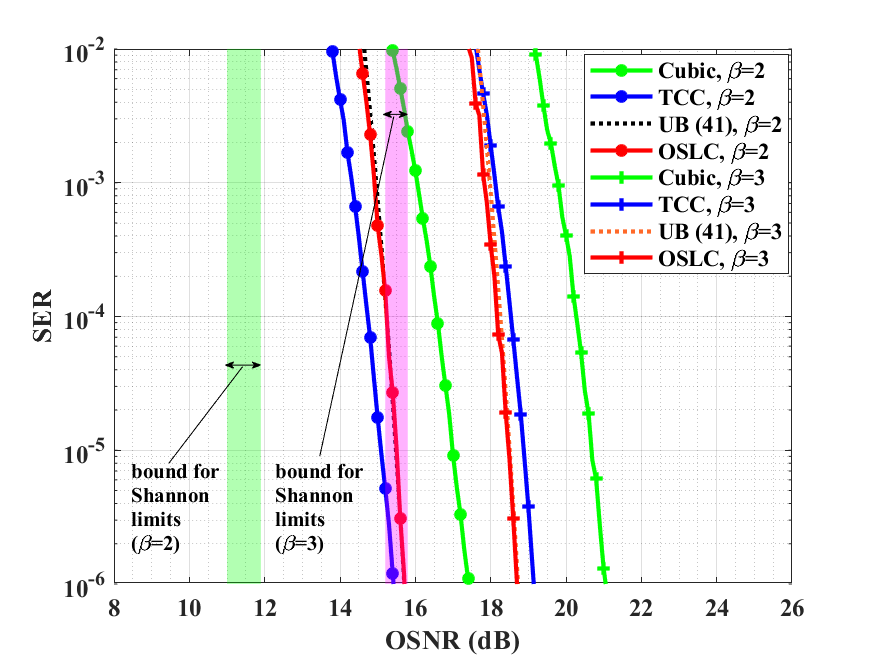}}
	\subfigure[The SER curves of the OSLC scheme and the benchmarks with $\EEE=0.3$ when $\beta=4$ bpcu and $5$ bpcu.]
	{\label{Fig.r3_beta45_SER}\includegraphics[width=0.45\textwidth]{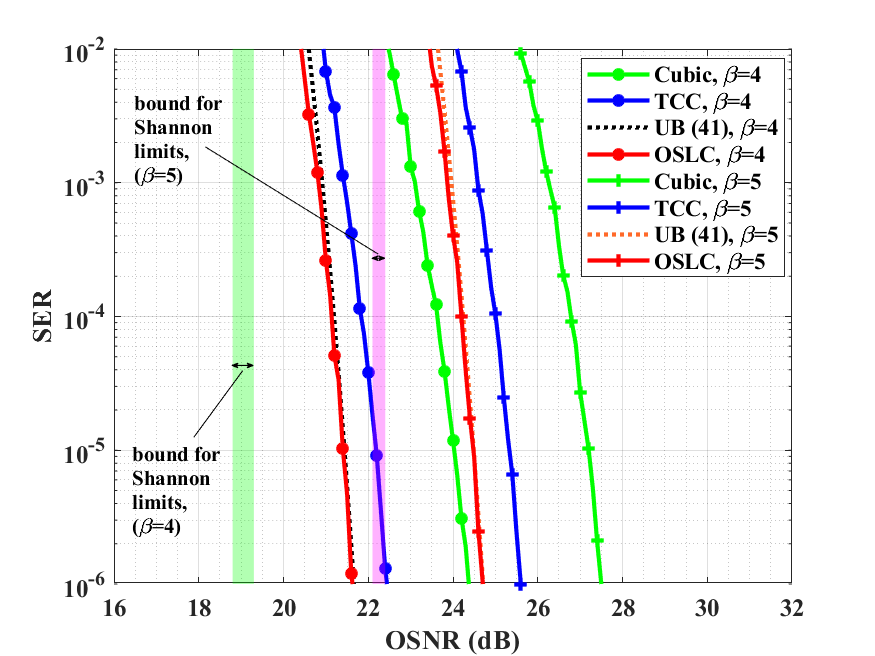}}
	\caption{The SER curves of the OSLC scheme and the benchmarks with $\EEE=0.3$.}
	\label{Fig.r3SER}
\end{figure*}

\subsubsection{AWGN Channels}
We first consider the SER performance of the above three constellations over the AWGN channel. By varying the normalized rate $\beta$ from $2$ bpcu to $5$ bpcu, Fig. \ref{Fig.r2SER} plots their SER curves with $\EEE=0.2$, while Fig. \ref{Fig.r3SER} plots those curves in the case of $\EEE=0.3$. 
%It can be seen that the OSLC significantly outperforms two benchmark schemes when the normalized rate $\beta=4$ and $5$. In the case of $\EEE=0.2$ and $\beta=4$ or $\beta=5$, the OSLC has OSNR gains of $2.3$ dB and $0.6$ dB at a target SER $10^{-5}$, as compared to the TCC and the cubic constellation, respectively, while OSNR gains of $2.1$ dB and $0.5$ dB can be observed when $\EEE=0.3$. 
It can be seen that the OSLC significantly outperforms two benchmark schemes when the normalized rate $\beta=4$ and $5$. In the case of $\EEE=0.2$ and $\beta=5$, the OSLC has OSNR gains of $3$ dB and $0.9$ dB at a target SER $10^{-5}$, as compared to  the cubic constellation and the TCC, respectively, while OSNR gains of $2.8$ dB and $0.9$ dB can be observed when $\EEE=0.3$. It is noted that  there is a $0.2$ dB difference in the OSNR gains compared with the cubic constellation (i.e., without constellation shaping), which is consistent with our result on the second-order asymptotics of the maximum shaping gain (see Fig. \ref{Fig.shaping_gain}). As the normalized rate $\beta$ decreases, the performance advantage of the OSLC over the TCC diminishes. From Fig. \ref{Fig.r2_beta23_SER} and Fig. \ref{Fig.r3_beta23_SER}, it can be observed that the SERs of the TCC are close to those of the OSLC at $\beta=3$, and at $\beta=2$ the TCC slightly outperforms the OSLC. The main reasons for this reduction in OSNR gains may come from three aspects: 1) by using the coarsely shaping and finely coding strategy, only a small part of the message sequence is used for geometrically shaping the OSLC when $\beta$ is small, which results to a shaping region of $\mathcal{L}_{24}$ that may be far less like the optimal shaping region; 2) the densely-packed structure \eqref{eq.OSLC} of the OSLC $\mathcal{L}_{24}$ may be broken in the regime of a small cardinality; 3) the lattice decoding algorithm leads to performance loss as compared with the ML decoding, especially when the cardinality of the lattice code is relatively small.

%The reduction in the SNR gain is primarily stemmed from the fact that the maximum shaping gain at any dimension is monotonically decreasing as the quantity $\EEE$ increases to $\frac{1}{2}$. 

It is also seen that the SERs of the OSLC can be well evaluated by 
\begin{flalign}\label{eq.UBE}
 P_{\text{e,UB}} = N_A Q\left( \kappa d_{\text{min}}/2\sigma\right), 
\end{flalign}
i.e., the union bound (UB) on the error probability for the ML decoding of the Leech lattice $\Lambda_{24}$, where $Q \left( u \right)\triangleq \dfrac1{\sqrt{2\pi}}\int_u^\infty e^{-y^2/2}dy$ is the Gaussian Q-function, and $N_A=196560$ and $d_{\text{min}}=4\sqrt{2}$ represent the kissing number and the MED of the Leech lattice $\Lambda_{24}$, respectively \cite{conwaysloane99_1}. This observation shows that the error performance of the bounded-distance algorithm used in our demodulator is very close to that of the best lattice decoding algorithm. In practical implementation, the UB \eqref{eq.UBE} can be used for predicting the error performance of the OSLC.

We also evaluate the performance gap between the OSLC and the Shannon limit. Note that there is no closed-form formula for the exact channel capacity of our considered VLC channel. Hence, we use the capacity upper and lower bounds given in \cite{Lapidoth2009Capacity,Farid2009Channel} and plot those bounds in Figs. \ref{Fig.r2SER} and \ref{Fig.r3SER} by semi-transparent strips. From Figs. \ref{Fig.r2_beta45_SER} and \ref{Fig.r3_beta45_SER}, it can be seen that, the gap between the OSLC and the ultimate limit are within 2.5 dB in the high-SNR regimes, which is mainly caused by the fundamental backoff due to finite blocklength.

%in the case of $\EEE=0.2$ and $\beta=4$, the gap between the OSLC and the Shannon limit is about $2.1-2.7$ dB at a target SER $10^{-5}$, and when $\beta=5$ the gap is $2.1-2.4$ dB. From Fig. , the gaps are about $2.1-2.6$ dB and $2.1-2.4$ dB in the case of $\EEE=0.3$ and $\beta=4$ and that of $\EEE=0.3$ and $\beta=5$, respectively.

\subsection{Indoor VLC Scenario}
We particularly provide a glimpse of the error performance of the OSLC in an indoor VLC scenario, where a $4$ m$\times4$ m$\times3$ m room with $4$ LED lamps of the same specification and a single receiver PD is considered. We assume that the LED lamps are installed on the ceiling and equipped with $7 \times 7$ LED array, and one receiver PD is located on the plane at height of $0.6$m. Other parameters of the LED transmitters and the PD receiver are listed in Tables \ref{Tab.transmitter} and \ref{Tab.receiver}, respectively. As pointed in \cite{Li2021Space}, if those LED lamps are required to have identical brightness, the considered MISO VLC channel is equivalent to a scalar VLC channel via spatial repetition coding, and our proposed constellation can be readily used. 

Denote the drive current of all LED chips by $I$. After removing the DC component and spatial repetition coding, the equivalent scalar VLC channel is given by 
\begin{flalign}\label{eq:indoorVLC1}
	R=\sum_{q=1}^{4}\sum_{j=1}^{49}  s\cdot h_{qj}\cdot \gamma \cdot (I-I_{\min}) + Z,
\end{flalign}
where the channel coefficients $h_{qj}$ are computed by Lambertian's model as in \cite{Komine2004Fundamental}. With the same treatment in \cite{Ma2015Coordinated}, the variance of the AWGN $\sigma^2$ is approximated by 
\begin{flalign}
    \sigma^2=2q_eB\left( \sum_{q=1}^{4}\sum_{j=1}^{49} s\cdot h_{qj} \cdot \gamma\cdot \mathbb{E}[I] +I_{\mathrm{bg}}I_2\right),
\end{flalign}
where the quantity $q_e$ denotes the electronic charge. It is straightforward that $I\in [0.4,0.6]$. Let $X=(I-I_{\min})/(I_{\max}-I_{\min})$ and the dimming factor $\alpha=(\mathbb{E}[I]-I_{\min})/(I_{\max}-I_{\min})$, and then, after an appropriate linear transform, the model \eqref{eq:indoorVLC1} can be rewritten as
\begin{flalign}\label{eq:indoorVLC2}
	R= X + \sigma^2/\left(0.2 s\cdot\gamma\cdot \left(\sum_{q=1}^{4}\sum_{j=1}^{49}   h_{qj} \right) \right),
\end{flalign}
where the block code $\mathcal{X}$ with blocklength $n$ for the new input $X$ is subject to the constraints \eqref{eq.constraint}.

\begin{figure}
	\centering
	\resizebox{8cm}{!}{\includegraphics[width=10cm]{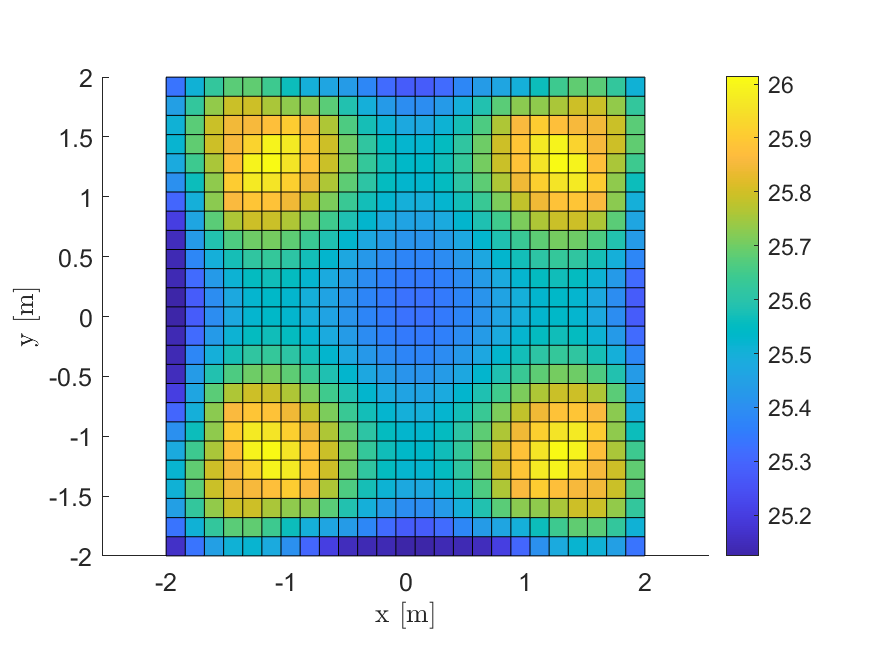}}\\
	\caption{The simulation of OSNR for the $4$m$\times4$m$\times3$m room.}
	\label{Fig.SNR}
\end{figure}

We first plot the simulated OSNR for the considered scenario in Fig. \ref{Fig.SNR}, from which it can be seen that the received OSNRs are about $25-26$ dB. In the last subsection, numerical results for the AWGN channel have indicated that high normalized rates can be achieved at those high SNRs. Table \ref{Tab.average SER} lists the average SER of our proposed scheme and the benchmarks with the normalized rate $\beta=5$ for the considered indoor channel, where the receiver PD is randomly located at $(X_P,Y_P)$ with two coordinates $X_P$ and $Y_P$ independently and uniformly distributed on the interval $[-4,4]$. Numerical results also verify the performance advantage of our schemes over others in the indoor VLC application.

\begin{table}[h]
	\caption{Transmitter Parameters}
	\centering
	\begin{tabular}{lr}%lcr分别为左对齐居中右对齐
		\toprule
		LED lamp coordinates & $\left(\pm1.6,\pm1.6,3\right)$ [m]\\
		\midrule
		Minimum drive current $I_{\min}$ & $0.4$ [A]\\
		\midrule
		Maximum drive current $I_{\max}$& $0.6$ [A]\\
		\midrule
		Semi-angle at half power & $60^{\circ}$\\
		\midrule
		Interval of the LED array & $1$ [cm]\\
		\midrule
		O/E conversion efficiency of LED $\gamma$ & $0.45$ [W/A]\\
		\bottomrule
	\end{tabular}

	\label{Tab.transmitter}
\end{table}

\begin{table}[h]
	\caption{Receiver Parameters}
	\centering
	\begin{tabular}{lr}%lcr分别为左对齐居中右对齐
		\toprule
		Physical area of PD & $1$ [cm$^2$]\\
		\midrule
		Gain of optical filter & $1$\\
		\midrule
		Refractive index of the lens at PD & $1.5$\\
		\midrule
		Responsitivity of PD $s$ & $0.4$ [A/W]\\
		\midrule
		Field of view (FOV) & $60^{\circ}$\\
		\midrule
		System bandwidth $B$ & $10$ [MHz]\\
		\midrule
		Noise bandwidth factor $I_{2}$ & $0.562$\\
		\midrule
		Background current $I_{bg}$ & $100$ [$\mu$A]\\
		\bottomrule
	\end{tabular}
	\label{Tab.receiver}
\end{table}

%\begin{tabular}{|c|c|c|}  
%	\hline  
%	1 & 1 & 3 \\  
%	\hline  
%	4 & 4 & 5 \\  
%	\hline  
%	6 & 6 & 7 \\  
%	\hline  
%\end{tabular}
%\begin{table}[h]
%	\caption{Average SER of the the $4$m$\times4$m$\times3$m room}
%	\centering
%	\begin{tabular}{|c|c|c|c|c|c|c|}
%		\hline
%		%\multirow{2}{*}{$\EEE$}& 
%		$\EEE$ & \multicolumn{3}{c}{0.2}& \multicolumn{3}{|c|}{0.3}\\
%		%\cline{1-7}   %在第2列到第5列下面划线
%		\hline
%		Scheme &Cubic &TCC &OSLC &Cubic &TCC &OSLC\\
%		\hline
%		Average SER &$0.0168$&$0.0013$&521&1922&d&d\\
%		\hline
%	\end{tabular}
%	\label{tab1}
%\end{table}

%\begin{table}[h]
%	\caption{Average SER of the the $4$m$\times4$m$\times3$m room}
%	\centering
%	\begin{tabular}{|c|c|c|}
%		\hline
%		%\multirow{2}{*}{$\EEE$}& 
%		$\EEE$ & Scheme & Average SER\\
%		%\cline{1-7}   %在第2列到第5列下面划线
%		\hline
%		\multirow{3}{*}{$0.2$} &Cubic constellation &$0.0168$ \\
%		\cline{2-3}
%		&TCC &$0.0013$ \\
%		\cline{2-3}
%		&OSLC &$7.501\times10^{-4}$ \\
%		\hline
%		\multirow{3}{*}{$0.3$} &Cubic constellation &$4.621\times10^{-4}$ \\
%		\cline{2-3}
%		&TCC &$1.065\times10^{-6}$ \\
%		\cline{2-3}
%		&OSLC &$\ll 10^{-6}$ \\
%		\hline
%	\end{tabular}
%	\label{Tab.average SER}
%\end{table}

\begin{table}[!h]
	\renewcommand{\arraystretch}{1}
	\caption{Average SER for the indoor VLC system.}
	\label{Tab.average SER}
	\centering
	\begin{tabular}{|c|c|c|}
		\hline
		\diagbox{Scheme}{Average SER}{Parameter $\EEE$} & $0.2$ & $0.3$ \\
		\hline
		Cubic constellation & $0.3004$ & $0.0196$  \\
		\hline
		TCC &$0.0241$ & $3.727\times10^{-6}$ \\
		\hline
		OSLC & $7.501\times10^{-4}$ & $\ll 10^{-6}$  \\
		\hline
	\end{tabular}
\end{table}
% \begin{table}[h]
% 	\caption{Receiver Parameters}
% 	\centering
%     \begin{tabular}{lr}%lcr分别为左对齐居中右对齐
% 		\toprule
% 		Physical area of PD & $1$ [cm$^2$]\\
% 		\midrule
% 		Gain of optical filter & $1$\\
% 		\midrule
% 		Refractive index of the lens at PD & $1.5$\\
% 		\midrule
% 		Responsitivity of PD & $0.4$ [A/W]\\
% 		\midrule
% 		FOV (field of view) & $60^{\circ}$\\
% 		\midrule
% 		System bandwidth & $10$ [MHz]\\
% 		\midrule
% 		Noise bandwidth factor & $0.562$\\
% 		\midrule
% 		Background current & $100$ [$\mu$A]\\
% 		\bottomrule
% 	\end{tabular}
% \label{Tab.receiver}
% \end{table}

%\rev{In \cite{Lapidoth2009Capacity}, the upper bound, lower bound, and the asymptotic capacity for the SISO-VLC were derived, which are solely determined by the \gjn{average-intensity constraint parameter $\EEE$.}} 
%As calculated in \cite[Theorem 3]{Lapidoth2009Capacity}, the corresponding SNR is $23.4-23.9$ dB when $\beta=5$ bpcu under the condition of $\EEE=0.2$, and the corresponding SNR is $21.9-22.3$ dB when $\beta=5$ bpcu under the condition of $\EEE=0.3$. Therefore, when $\beta=5$ bpcu, the difference between the channel capacity and the OSLC scheme is $2.1-2.6$ dB with $\EEE=0.2$, and $2.2-2.6$ dB with $\EEE=0.3$. \gjn{We can also conslude that under the SNR condition we simulated in Fig. \ref{Fig.SNR}, the proposed OSLC scheme has better error performance than the existing schemes.}

\section{Conclusion}\label{Sec.conclusion}
In this paper, a general framework for signaling over the short-packet VLC channel with a peak- and an average-intensity constraints is proposed. It mainly consists of finite blocklength analysis of geometric shaping and an elaborated method of constructing geometrically-shaped multi-dimensional constellations. Additionally, we present a 24-dimensional constellation design based on the Leech lattice as an illustrative example, which shows a significant OSNR gain as compared with existing schemes.

\section*{Acknowledgement} %Ru-Han Chen wishes to thank Prof. Iosif Pinelis and Prof. Guo-En Hu for their helpful comments.

This work is supported by the National Natural Science Foundation of China (NSFC) under Grant (62071489). 

\appendices

\section{Large Deviation of Sum of Independent Random Variables Uniformly Distributed on $[0,1]$}\label{app.ldt}
Let the random variable $U$ be uniformly distributed on $[0,1]$. For $s>0$, the moment-generating function of $U$ is
$$
M(s)\triangleq \mathbb{E}\left[\exp\left( s U \right) \right]=\frac{\exp(s)-1}{s},
$$
and we define
$$	
K(s)\triangleq\frac{M'(s)}{M(s)}=\frac{\exp(s)}{\exp(s)-1}-\frac{1}{s}.
$$
It is clear that $K(s)$ is monotone increasing and satisfies $K(0^{+})=\frac{1}{2}$ and $K(+\infty)=1$. Then, we let $s_x$ denote the unique solution to the equation 
\begin{flalign}
K(s_x)=x,
\end{flalign}
where $x\in[\frac{1}{2},1]$. Specially, an immediate consequence of \eqref{eq.solution} is 
\begin{flalign}
{\mu}^{*}=s_{1-\EEE}.	\label{eq.26}
\end{flalign}

Clearly, the solution $s_x$ increases with $x$. Then, for $x\in \left[\frac{1}{2},1\right]$, we define 
$$
L(x)\triangleq \log \left( M(s_x) \right)-xs_x,
$$
and
$$
D(x)\triangleq s_x \sqrt{K'(s_x)}=\sqrt{\frac{-s_x^2\cdot\exp(s_x)}{\left(
		\exp(s_x)-1	\right)^2
	}+1}.
$$
It can be easily verified both $L(x)$ and $D(x)$ are smooth. Then, by computing high-order derivatives of $D(x)$, we can show that $D(x)$ is bounded and non-decreasing with $x$. Also, we note that
\begin{flalign}
L{'}(x)&=\left(\frac{M{'}(s_x)}{M(s_x)}-x\right)\cdot \frac{\dd s_x}{\dd x}-s_x\\
&=\left(K\left(s_x\right)-x\right)\cdot \frac{\dd s_x}{\dd x}-s_x\\
&=-s_x.
\end{flalign}
Since $s_x$ is increasing as $x$ increases, the function $L(x)$ is concave and thus satisfies
\begin{flalign}
L(x)\le L(1-\tau)-s_{1-\tau}\left(x-(1-\tau)\right).
\end{flalign}
for any $x$ and $\tau$ in $[\frac{1}{2},1]$.

Denote the tail probability of the sample mean $\bar{U}_n=\frac{1}{n}\sum_{i=1}^n U_i$ by $\GG_n(x)=\mathbb{P}\left\{  \bar{U}_n \ge x \right\}$. Based on Theorem 1 in \cite{Petrov1965LDT}, the sequence of functions
%  Now we focus on the function series $\left\{\frac{G_n(s)}{G_n(y)}\right\}$.
\begin{flalign}\label{eq.large_deviation}
\GG_n(x)=\frac{\exp\left(n L(x)\right)}{  \sqrt{2\pi n} D(x) }	\left(1+o\left(1\right)\right)
\end{flalign}
as $n\to \infty$ uniformly in $x$ in the closed interval $\left[\frac{1}{2}+\epsilon,1-\epsilon\right]$, where $\epsilon$ is an arbitrarily positive constant.

\vspace{0.5cm}
\section{Proof of Theorem~\ref{thm.convergence}}
\label{app.convergence}
The average first moment of $\mathscr{T}_n(n\tau)$ can be alternatively expressed by the conditional expectation of the Irwin-Hall distribution as follows 
\begin{flalign}
\PP_n(n\tau)&=\frac{1}{n}\sum_{i=1}^n \mathbb{E}\left[ U_i\bigg\vert \sum_{i=1}^nU_i\le n\tau \right]\\
&=\mathbb{E}\left[ \bar{U}_n \Big\vert \bar{U}_n \le \tau  \right]\\
&=1-\mathbb{E}\left[ \bar{U}_n \Big\vert \bar{U}_n \ge 1-\tau  \right].\\
%&=\rev{\psi_n(n\tau)}
\end{flalign}
Remind that the tail probability $\GG_n(x)=\mathbb{P}\left\{  \bar{U}_n \ge x \right\}$. Then, it is straightforward that
\begin{flalign}
\PP_n(n\tau)&= 1+\frac{\int_{1-\tau}^{1}  x \dd \GG_n}{\GG_n(1-\tau)}\\
&=\tau-\int_{1-\tau}^{1}\frac{\GG_n(x)}{\GG_n(1-\tau)}\dd x.
\end{flalign}
The first-moment condition \eqref{eq.shaping parameter} immediately shows that
\begin{flalign}\label{eq.44}
\EEE=\tau_n^{\star}-\int_{1-\tau_n^{\star}}^{1}\frac{\GG_n(x)}{\GG_n(1-\tau_n^{\star})}\dd x.
\end{flalign}
It is clear that $\tau_n^{\star}\ge \EEE$ for any $n$. Next, we will prove that
\begin{flalign}
\lim_{n\to \infty}n\left( \tau_n^{\star}-\EEE \right)=\frac{1}{s_{1-\EEE}}.
\end{flalign}

Note that $\tau\ge \EEE$ and $\PP_n(1)=\frac{1}{2}$. For any $\tau\in \left( 0,1\right)$, there exists a positive (and possibly arbitrarily small) constant $\epsilon$ such that $1-\tau+2\epsilon\le 1$. Then we note that
\begin{flalign}
0\le \int_{1-\tau}^{1}\frac{\GG_n(x)}{\GG_n(1-\tau )}\dd x\le 2\epsilon +\int_{1-\tau+\epsilon}^{1-\epsilon}\frac{\GG_n(x)}{\GG_n(1-\tau )}\dd x.
\end{flalign}
Due to the uniform convergence of $\GG_n(x)$, we can rewrite 
\begin{flalign}
&\int_{1-\tau+\epsilon}^{1-\epsilon}{\GG_n(x)}\dd x\\ \nonumber
=&\int_{1-\tau+\epsilon}^{1-\epsilon} \frac{\exp\left(n L(x)\right)}{  \sqrt{2\pi n} D(x) }	\left(1+o\left(1\right)\right)\dd x  \\ \nonumber
\le &\int_{1-\tau+\epsilon}^{1-\epsilon} \frac{\exp\left(n \left(  
	L(1-\tau)-s_{1-\tau}\left(x-(1-\tau)\right)
	\right)	\right)}{  \sqrt{2\pi n} D(1-\tau) }	\left(1+o\left(1\right)\right)\dd x \\ \nonumber
\le&\frac{1}{s_{1-\tau}}\GG_n(1-\tau )\left(\frac{1}{n}+o\left( \frac{1}{n}\right)\right)
\end{flalign}

Immediately, we have $\tau_n^{\star}=\EEE+o(1)$. Note that 
\begin{flalign}
&n\left( \tau_n^{\star}-\EEE \right) \\ \nonumber
=&n\int_{1-\tau_n^{\star}}^{1}\frac{\GG_n(x)}{\GG_n(1-\tau_n^{\star})}\dd x \\
\le& n\left(\int_{1-\tau_n^{\star}}^{1-\epsilon}\frac{\GG_n(x)}{\GG_n(1-\tau_n^{\star})}\dd x +\epsilon
\frac{\GG_n(1-\epsilon)}{\GG_n(1-\tau_n^{\star})}
\right)  \\
=&\frac{1}{s_{1-\EEE}}+o(1)\label{eq.43}
\end{flalign}
On the one hand, for any finite $k\in \mathbb{N}$, we have 
\begin{flalign}
&n\int_{1-\tau_n^{\star}}^{1}\frac{\GG_n(x)}{\GG_n(1-\tau_n^{\star})}\dd x  \\ \nonumber
\ge&  n\int_{1-\tau_n^{\star}}^{1-\tau_n^{\star}+\frac{k}{n}}\frac{\GG_n(x)}{\GG_n(1-\tau_n^{\star})}\dd x \\
=& n\int_{0}^{\frac{k}{n}}
\exp\left(  -ns_{1-\tau_n^{\star}}x 
\right)\dd x  \cdot(1+o(1))\\
=&\frac{1-\exp\left(  -ks_{1-\tau_n^{\star}} 
	\right)}{s_{1-\EEE}}(1+o(1)).\label{eq.46}
\end{flalign}
\eqref{eq.26}, \eqref{eq.43} and \eqref{eq.46} complete the proof of Theorem \ref{thm.convergence}.

\section{Proof of Theorem~\ref{thm.shaping gain}}
\label{app.shaping gain}

It should be noted that:
\begin{flalign}
&\log\left(	\overline{\mathtt{SG}}_{\textnormal{VLC}}(n;\alpha) \right) +\log\left({2\EEE}\right) \nonumber \\
=& \frac{1}{n}\log\left({\rm vol}\left(\mathscr{T}_n \left(t_n^{\star}\right)\right)\right) \nonumber \\
=& \frac{1}{n}\log\left(\GG_n \left( 1-\tau_n^{\star} \right)\right) \nonumber \\
=&  \frac{1}{n}\log\left(\frac{\exp\left(n L\left(1-\tau_n^{\star}\right)\right)}{  \sqrt{2\pi n} D\left(1-\tau_n^{\star}\right) }	\left(1+o\left(1\right)\right) \right) \label{eq.shaping gain-1}\\
=&  L\left(1-\EEE-\frac{1}{\mu^*}\frac{1}{n}+o\left(\frac{1}{n}\right)\right)+o\left(\frac{1}{n}\right)\nonumber\\
&-\frac{1}{n}\log\left(  \sqrt{2\pi n}D\left(1-\EEE-\frac{1}{\mu^*}\frac{1}{n}+o\left(\frac{1}{n}\right)\right)\right)  \IEEEeqnarraynumspace \label{eq.shaping gain-2}  \\
=&L\left(1-\EEE\right) -\frac{\log(2\pi n)}{2n} +\frac{1}{n} \left(1-\log \left( D\left( 1-\alpha \right) \right) \right)+o\left( \frac{1}{n}\right)
	\label{eq.shaping gain-3}\\
=&\hh_{\text{max}}(\alpha) -\frac{\log(n)}{2n} +o\left( \frac{1}{n}\right) \nonumber\\
&+\frac{1}{n}\left(
	1-\frac{1}{2}\log\left( 2\pi\left( -\mu^*+2\alpha \mu^*+\left(\mu^* \right)^2\alpha\left(1-\alpha\right)\right) \right)
	\right) \IEEEeqnarraynumspace \label{eq.shaping gain-4}\\
=&\hh_{\text{max}}(\alpha)-\log(2\EEE)-\frac{\log(n)}{2n} +\frac{\omega_{\alpha}}{n}
+o\left(\frac{1}{n}\right)
\end{flalign}
where Eq. \eqref{eq.shaping gain-1} is obtained by Eq. \eqref{eq.large_deviation}, Eq. \eqref{eq.shaping gain-2} is derived from Theorem \ref{thm.convergence} Eq. \eqref{eq.shaping gain-3} is calculated by
\begin{flalign}
	&L\left(1-\EEE-\frac{1}{\mu^*}\frac{1}{n}+o\left(\frac{1}{n}\right)\right)  \nonumber \\
	=& L\left(1-\EEE\right) +L'\left(1-\EEE\right)\left(-\frac{1}{\mu^*}\frac{1}{n}+o\left(\frac{1}{n}\right)\right)+o\left(\frac{1}{n} \right)  \nonumber \\
	=&L\left(1-\EEE\right)-{s_{1-\EEE}}\left( -\frac{1}{\mu^*}\frac{1}{n} \right)+o\left(\frac{1}{n}\right)\nonumber \\
	=&L\left(1-\EEE\right)+\frac{1}{n} +o\left(\frac{1}{n}\right),
\end{flalign}
and $D\left(1-\EEE-\frac{1}{\mu^*}\frac{1}{n}+o\left(\frac{1}{n}\right)\right)=D\left(1-\EEE\right)+o\left(1\right)$, and Eq. \eqref{eq.shaping gain-4} can be obtained by substitute the value into $L\left(1-\EEE\right)$ and $D\left( 1-\alpha \right)$.

\bibliographystyle{ieeetr}
\bibliography{biblio1}

%%%%%%%%%%%%%%%%%%%%%%%%%%%%%%%%%%%%%%%%%%%%%%%%%%%%%%%%%%%%%%%%%%%%%
%\addtolength{\textheight}{-68mm}
%%%%%%%%%%%%%%%%%%%%%%%%%%%%%%%%%%%%%%%%%%%%%%%%%%%%%%%%%%%%%%%%%%%%%

\end{document}